\documentclass[aps,pre,preprint,notitlepage,superscriptaddress]{revtex4-2}
\pdfoutput=1
\usepackage{amsmath}      % 数学公式增强（gather、align、eqref等）
\usepackage{amsfonts}     % 数学字体（\mathbb、\mathbf等）
\usepackage{bm}           % 粗体数学符号（\bm）
\usepackage{graphicx}     % 插图功能
\usepackage[hidelinks]{hyperref}   % 超链接（引用跳转、DOI链接等）
\usepackage{subcaption}
\usepackage{lineno}
\usepackage{comment}
\usepackage{placeins}
\usepackage{float}

\begin{document}
	%\linenumbers
\raggedbottom
	
	%\preprint{}
	
	\title{Multiple-relaxation-time total enthalpy lattice Boltzmann model for liquid–vapor phase change}
	
	
	%\author{Author Name}
	%\thanks{}
	%\affiliation{Affiliation}
	
	\author{Hangyu He}
	\affiliation{School of Mathematics and Statistics, Changsha University of Science and Technology, Changsha 410114, China}
	\affiliation{Hunan Provincial Key Laboratory of Mathematical Modeling and Analysis in Engineering, Changsha University of Science and Technology, Changsha 410114, China}
	
%	\author{Rongzong Huang}
%	\affiliation{School of Energy Science and Engineering, Central South University, Changsha 410083, China}
	
	\author{Lei Wang}
	\affiliation{School of Mathematics and Physics, Center for Mathematical Sciences, China University of Geosciences, Wuhan 430074, China}
		
	\author{Xuguang Yang}
	\affiliation{School of Mathematics and Statistics, Hunan First Normal University, Changsha 410205, China}
	
	\author{Yong Zhao}
	\email{zhaoy@csust.edu.cn}
	\affiliation{School of Mathematics and Statistics, Changsha University of Science and Technology, Changsha 410114, China}
	\affiliation{Hunan Provincial Key Laboratory of Mathematical Modeling and Analysis in Engineering, Changsha University of Science and Technology, Changsha 410114, China}
	\affiliation{Hubei Key Laboratory of Engineering Modeling and Scientific Computing, Huazhong University of Science Changsha University of Science and Technology, Wuhan 430074, Hubei, China}
	
	
	%\date{}
	
	\begin{abstract}
	In this paper, a multiple-relaxation-time (MRT) total enthalpy lattice Boltzmann (LB) model is proposed for liquid-vapor phase change. The pseudopotential LB method is employed to describe the liquid-vapor multiphase dynamics, while the thermal field is solved using an MRT LB formulation with total enthalpy as the evolution variable, thereby incorporating sensible and latent heat into a unified framework. A latent heat expression based on the equation of state (EOS) is further derived to provide a thermodynamic reference for its specification. Multiscale analysis demonstrates that the proposed model recovers the target macroscopic energy equation without introducing the nonlinear thermodynamic source term commonly required in temperature-based formulations, thereby avoiding the explicit evaluation of $\nabla\cdot\mathbf{u}$. The model is validated through free droplet evaporation, droplet evaporation on a heated surface, and bubble growth and departure during boiling, showing good agreement with published results. Simulations with different prescribed latent heats are further conducted in all three cases, demonstrating that the latent heat can be independently specified according to the physical conditions considered without altering the  equation of state.
	\end{abstract}

	%\keywords{}
	
	\maketitle
	
	\section{INTRODUCTION}	
	\label{sec:INTRODUCTION}	
	Liquid-vapor phase change is accompanied by substantial latent heat absorption or release, enabling highly efficient energy transport under relatively small temperature differences~\cite{Wilson2023,Cho2017}. Owing to this distinctive characteristic, phase-change heat transfer has been widely utilized in energy and power systems, refrigeration and air-conditioning technologies, electronic thermal management, and other high-heat-flux applications~\cite{Cho2017,Kharangate2017,Upot2023}. However, liquid-vapor phase change processes involve strongly coupled phenomena, including interfacial evolution~\cite{Sato2013}, phase equilibrium and thermodynamic consistency~\cite{Faccanoni2012}, as well as interfacial heat and mass transfer~\cite{Wenzel2023}, which makes a comprehensive understanding and accurate prediction of the underlying transport mechanisms challenging. Consequently, the development of accurate and efficient numerical methods for liquid-vapor phase change has attracted considerable attention~\cite{Li2016,Liu2015}.
			
	The lattice Boltzmann method (LBM) is a mesoscopic numerical approach rooted in kinetic theory, in which the evolution of particle distribution functions on a discrete lattice is employed to recover the macroscopic governing equations of fluid flow~\cite{McNamara1988,Chen1998,Succi2001,Aidun2010}. Compared with conventional numerical methods based directly on the discretization of the Navier--Stokes equations, the LBM features a simple collision--streaming algorithm, intrinsic locality, straightforward treatment of complex boundaries, and high parallel efficiency~\cite{Chen1998,Aidun2010}. Owing to these characteristics, the LBM has been extensively applied to complex fluid systems, particularly multiphase flows~\cite{Gunstensen1991} and thermal flows~\cite{He1998}. Recent developments have further extended the LB framework to multicomponent flows through modified pseudopotential formulations~\cite{Zhao2021Multicomponent} and to conjugate heat-transfer problems using MRT schemes~\cite{Zhao2025CHT}. The LBM has also been widely employed for simulations of liquid--vapor phase change heat transfer~\cite{Li2016,Li2017,Huang2021}.
	
    In particular, the pseudopotential method has been widely used for
    liquid-vapor phase change simulations~\cite{Gong2012,Zhang2021,Wang2022}.
    In this method, the interactions between neighboring fluid particles are
    represented through attractive or repulsive forces, thereby introducing
    nonideal fluid behavior into the LB framework~\cite{Shan1993,Li2013}.
    Consequently, liquid-vapor phase separation and interface evolution can
    occur naturally without explicitly tracking the phase interface~\cite{Shan1993,Gong2012}. For thermal pseudopotential models, the macroscopic temperature equation commonly employed for liquid-vapor phase change can be derived from the local entropy balance~\cite{He1998,Gong2012}
    \begin{equation}
    \rho T\frac{Ds}{Dt}=\nabla\cdot(\lambda\nabla T),
    \label{eq:entropy_balance}
    \end{equation}
    together with the thermodynamic relation for a nonideal fluid. The resulting temperature equation can be written as
    \begin{equation}
    \rho c_v\left(\frac{\partial T}{\partial t}+\mathbf{u}\cdot\nabla T\right)
    =\nabla\cdot(\lambda\nabla T) -T\left(\frac{\partial p_{\rm EOS}}{\partial T}\right)_{\rho}\nabla\cdot\mathbf{u}.
    \label{eq:temperature_entropy}
    \end{equation}
    Accordingly, many thermal pseudopotential LB models~\cite{Gong2012,Li2017,Zhang2021} have been developed to
    recover this temperature equation or its modified forms.
    
    An early representative model was proposed by Gong and Cheng~\cite{Gong2012}, in which the pseudopotential multiphase model was coupled with a temperature-based energy equation to simulate liquid-vapor phase change processes. Li et al.~\cite{Li2017} subsequently improved the thermal formulation by correcting the treatment of the thermal diffusion term and eliminating additional error terms in the recovered temperature equation. They~\cite{Li2022} later extended the framework to three dimensions using a compact D3Q7 lattice. However, the gradient of volumetric heat capacity $\nabla(\rho c_v)$ was still involved in the formulation. More recently, Wang et al.~\cite{Wang2022} developed a compact D2Q5 thermal LB model that avoids the explicit evaluations of $\nabla\cdot(\lambda\nabla T)$ and $\nabla(\rho c_v)$; nevertheless, the recovered temperature equation still contains a thermodynamic source term involving the velocity divergence $\nabla\cdot\mathbf{u}$. Of particular relevance to the present work, Zhang et al.~\cite{Zhang2021} proposed an MRT thermal LB model in which the latent heat of vaporization was decoupled from the equation of state and could therefore be independently specified. Beyond these temperature-based thermal LB formulations, Qiao et al.~\cite{Qiao2026} developed a thermodynamically consistent non-isothermal two phase LB model based on the Onsager variational principle and a total energy formulation.
    
    Nevertheless, within pseudopotential LB simulations of liquid-vapor phase
    change, temperature-based thermal formulations remain predominant. the above liquid-vapor phase change formulations remain essentially temperature based, with phase change thermodynamics represented through the temperature equation and associated source terms. Total enthalpy formulations have also been developed within the LB framework for solid-liquid phase change problems, in which sensible and latent heat are incorporated into a unified enthalpy variable~\cite{Huang2013SolidLiquid}. In contrast, the present work develops an MRT total enthalpy LB formulation for liquid-vapor phase change, in which the pseudopotential method is employed for multiphase dynamics and sensible and latent heat are incorporated into the evolution variable through a unified total enthalpy formulation. The target macroscopic total enthalpy equation is recovered without introducing the nonlinear thermodynamic source term, thereby avoiding the explicit evaluation of $\nabla\cdot\mathbf{u}$ while retaining the local nature of the LB scheme. The formulation further permits flexible specification of caloric properties. In addition, a separate EOS-based expression for evaluating the latent heat is derived to provide a thermodynamic reference for its specification.
    
    The rest of the paper is organized as follows. In Sec.~\ref{sec:IMPROVED THERMALMRT LB MODEL}, the proposed total enthalpy thermal LB model is presented, together with the derivation of an EOS-based latent heat expression and a multiscale analysis of the recovered macroscopic energy equation. In Sec.~\ref{sec:PSEUDOPOTENTIAL LB MODEL}, the pseudopotential multiphase LB model is introduced. Numerical validations and discussions are provided in Sec.~\ref{sec:RESULTS AND DISCUSSION}. Finally, the main conclusions are summarized in Sec.~\ref{sec:CONCLUSION}.
    
    \section{IMPROVED THERMALMRT LB MODEL}
    \label{sec:IMPROVED THERMALMRT LB MODEL}
    
    \subsection{Macroscopic energy equation and total enthalpy formulation}  
    To incorporate the latent heat effects during liquid-vapor phase change, we begin with the general energy conservation equation~\cite{Welch2000,Hardt2008}. Neglecting viscous dissipation and external heat sources, the energy equation for a compressible fluid reads.
    \begin{equation}
    \partial_t (\rho e) + \nabla \cdot (\rho \mathbf{u} e) =  - p \nabla \cdot \mathbf{u}+\nabla \cdot(\lambda \nabla T) ,\label{eq:energy_general}
    \end{equation}
    where $e$ denotes the specific internal energy. Introducing the specific enthalpy $h=e+p/\rho$ transforms Eq.~\eqref{eq:energy_general} into its enthalpy form, following the enthalpy formulation widely used for phase change heat transfer~\cite{Voller1987,Huang2016SolidLiquid}.
    \begin{equation}
    \partial_t (\rho h) + \nabla \cdot (\rho \mathbf{u} h) = \partial_t p+ \mathbf{u} \cdot \nabla p+\nabla \cdot(\lambda \nabla T),\label{eq:energy_general_transforms}
    \end{equation}
    In the present thermal formulation, a reduced caloric closure is adopted.
    The explicit energetic contribution associated with the material pressure
    variation, represented by $Dp/Dt=\partial_t p+ \mathbf{u} \cdot \nabla p$ in Eq.~\eqref{eq:energy_general_transforms}, is not retained, whereas the
    sensible and latent enthalpy contributions are explicitly accounted for.
    Such a treatment is consistent with liquid-vapor phase change models in
    which pressure is retained in the hydrodynamic and interfacial mechanical
    description, while the thermal process is described through temperature
    or enthalpy together with latent heat ~\cite{Hardt2008,Badillo2012,Cipriano2024}.
    Pressure-work simplifications have also been adopted in thermal LB and
    low-Mach evaporation models~\cite{Yang2018,Cipriano2024,Zhao2023}.
    It should be emphasized that this approximation does not imply
    $Dp/Dt=0$, but only neglects its explicit contribution to the caloric
    energy balance. Accordingly, Eq.~\eqref{eq:energy_general_transforms}
    reduces to
    \begin{equation}
    \partial_t (\rho h) + \nabla \cdot (\rho \mathbf{u} h) = \nabla \cdot (\lambda \nabla T).
    \label{eq:enthalpy_simplified}
    \end{equation}
    
    To close Eq.~\eqref{eq:enthalpy_simplified}, a constitutive relation between the specific enthalpy and temperature is required. Taking $T_{\rm ref}$ as a prescribed reference temperature, the specific enthalpies of the liquid and vapor phases are defined as~\cite{Chakraborty2007}
    \begin{equation}
    h_l=c_{p,l}(T-T_{\rm ref}),
    \label{eq:app_hl}
    \end{equation}
    and
    \begin{equation}
    h_v=c_{p,v}(T-T_{\rm ref})+L,
    \label{eq:app_hv}
    \end{equation}
    respectively, where $c_{p,l}$ and $c_{p,v}$ are the specific heat capacities of the liquid and vapor phases, and $L$ denotes the latent heat of vaporization.
    
    Within the diffuse liquid-vapor interface, the local phase volume
    fractions can be related to the density~\cite{MohammadiShad2017}. The
    liquid volume fraction is evaluated as
    \begin{equation}
    f_l = \frac{\rho-\rho_v} {\rho_l-\rho_v},
    \label{eq:app_fl}
    \end{equation}
    while
    \begin{equation}
    f_v=1-f_l.
    \label{eq:app_fv}
    \end{equation}
    Accordingly, the local mixture density satisfies
    \begin{equation}
    \rho=f_l\rho_l+f_v\rho_v.
    \label{eq:app_rho_mix}
    \end{equation}
    
    Since $h_l$ and $h_v$ are specific enthalpies per unit mass, the mixture enthalpy should be weighted by the corresponding mass fractions. The liquid and vapor mass fractions are therefore defined as
    \begin{equation}
    Y_l=\frac{f_l\rho_l}{\rho},
    \qquad
    Y_v=\frac{f_v\rho_v}{\rho},
    \label{eq:app_mass_fraction}
    \end{equation}
    which satisfy
    \begin{equation}
    Y_l+Y_v=1.
    \end{equation}
    
    The mixture specific enthalpy is then expressed as
    \begin{equation}
    h=Y_lh_l+Y_vh_v.
    \label{eq:app_hmix}
    \end{equation}
    To formulate the thermal energy equation in terms of a volumetric
    quantity, the volumetric total enthalpy, denoted by $H$, is introduced,
    following the enthalpy based lattice Boltzmann formulation for
    phase change problems~\cite{Huang2016SolidLiquid}.
    \begin{equation}
    H \equiv \rho h.
    \label{eq:H_definition}
    \end{equation}
    Substituting Eq.~\eqref{eq:app_mass_fraction} into 
    Eq.~\eqref{eq:app_hmix} and using Eq.~\eqref{eq:H_definition} yields
    \begin{equation}
    H=f_l\rho_lh_l+f_v\rho_vh_v.
    \label{eq:app_Hmix}
    \end{equation}
    Using Eqs.~\eqref{eq:app_hl} and \eqref{eq:app_hv}, one obtains
    \begin{align}
    H ={}&f_l\rho_lc_{p,l}(T-T_{\rm ref})\nonumber\\
    &+ f_v\rho_v \left[ c_{p,v}(T-T_{\rm ref})+L\right],
    \end{align}
    which can be rearranged as
    \begin{equation}
    H =\left( f_l\rho_lc_{p,l} + f_v\rho_vc_{p,v}\right) (T-T_{\rm ref})+ f_v\rho_vL.
    \label{eq:app_H_final1}
    \end{equation}
    For convenience, the effective volumetric heat capacity of the mixture is defined as
    \begin{equation}
    (\rho c_p)_m = f_l\rho_lc_{p,l} + f_v\rho_vc_{p,v}.
    \label{eq:app_rhocp}
    \end{equation}
    Therefore,
    \begin{equation}
    H	=	(\rho c_p)_m(T-T_{\rm ref})	+f_v\rho_vL,
    \label{eq:app_H_final}
    \end{equation}
    and the temperature can be recovered from the total enthalpy through
    \begin{equation}
    T	=	T_{\rm ref}	+\frac{	H-f_v\rho_vL}{(\rho c_p)_m}.
    \label{eq:app_T_recovery}
    \end{equation}
    With the definition $H=\rho h$, the macroscopic enthalpy equation,
    Eq.~\eqref{eq:enthalpy_simplified}, can be rewritten in terms of the
    volumetric total enthalpy as
    \begin{equation}
    \partial_t H+\nabla\cdot(H\mathbf{u})=\nabla\cdot(\lambda\nabla T).
    \label{eq:enthalpy_H}
    \end{equation}
      
    Therefore, the above derivation provides a unified total enthalpy
    representation for the liquid-vapor mixture, in which sensible and latent
    heat are incorporated into a single thermal variable, following the general enthalpy-based treatment of phase change heat transfer~\cite{Voller1987,Chakraborty2007}. The resulting enthalpy-temperature relation and macroscopic total enthalpy
    equation form the thermodynamic basis of the present thermal LB model.
    In the present work, the total enthalpy formulation is coupled with the
    pseudopotential LB framework to describe the hydrodynamic and thermal
    behaviors during liquid-vapor phase change. The thermal LB model is formulated directly within the MRT framework. The MRT approach has been widely employed in thermal LB models for liquid-vapor phase change simulations~\cite{Li2017,Zhang2021}.
    
    \subsection{Thermodynamic evaluation of the latent heat}
    
   In the present total enthalpy formulation, the latent heat $L$ is treated
   as an independently specifiable caloric property and is therefore not
   constrained by the equation of state. Nevertheless, for a given EOS, a
   thermodynamically consistent latent heat can be evaluated from the
   liquid-vapor coexistence properties and used as a reference for the
   specification of $L$~\cite{Biferale2013}. In this subsection, such an EOS-based latent heat is derived from the Clausius-Clapeyron relation~\cite{Callen1985}.
   
    \begin{equation}
    L=T\left(\frac{1}{\rho_v}- \frac{1}{\rho_l}\right) \frac{d p_{\rm sat}}{dT},
    \label{eq:app_clapeyron}
    \end{equation}
    where $\rho_l$, $\rho_v$, and $p_{\rm sat}$ are the saturated liquid density, saturated vapor density, and saturation pressure,respectively. For a thermodynamically consistent liquid vapor coexistence state,the Maxwell equal area condition can be expressed as~\cite{Yuan2006}
    \begin{equation}
    \int_{v_l}^{v_v}
    \left[
    p(v,T)-p_{\rm sat}(T)
    \right]\,dv=0,
    \label{eq:app_maxwell}
    \end{equation}
    where $v_l=1/\rho_l$ and $v_v=1/\rho_v$ denote the specific volumes of the saturated liquid and vapor phases, respectively. Differentiating Eq.~\eqref{eq:app_maxwell} with respect to
    temperature gives
    \begin{equation}
    (v_v-v_l)\frac{dp_{\rm sat}}{dT}
    =\int_{v_l}^{v_v} \left(\frac{\partial p}{\partial T}\right)_vdv.
    \label{eq:app_dpsat}
    \end{equation}
    Substituting Eq.~\eqref{eq:app_dpsat} into Eq.~\eqref{eq:app_clapeyron} yields
    \begin{equation}
    L=T \int_{v_l}^{v_v}\left(\frac{\partial p}{\partial T}\right)_vdv.
    \label{eq:app_latent_integral}
    \end{equation}
    
    The PR EOS employed in the present model is written as~\cite{PengRobinson1976}
    \begin{equation}
    p =\frac{RT}{v-b}-\frac{a\alpha(T)}{v^2+2bv-b^2},
    \label{eq:PR_EOS_simplified}
    \end{equation}
    where $R$ is the gas constant and $a$ and $b$ are the PR EOS parameters,
    given by
    \begin{equation}
    a=0.45724\frac{R^2T_c^2}{p_c},\qquad 
    b=0.07780\frac{RT_c}{p_c},
    \label{eq:PR_ab}
    \end{equation}
    with $T_c$ and $p_c$ denoting the critical temperature and pressure,
    respectively. The temperature-dependent attraction function is
    \begin{equation}
    \alpha(T)=\left[1+\kappa\left(1-\sqrt{T_r}\right)\right]^2,
    \qquad
    T_r=\frac{T}{T_c},
    \label{eq:PR_alpha}
    \end{equation}
    where
    \begin{equation}
    \kappa=0.37464+1.54226\omega-0.26992\omega^2,
    \label{eq:PR_kappa}
    \end{equation}
    and $\omega$ is the acentric factor.
    
    At a fixed specific volume $v$, differentiation of
    Eq.~\eqref{eq:PR_EOS_simplified} with respect to temperature gives
    \begin{equation}
    \left(\frac{\partial p}{\partial T}\right)_v=\frac{R}{v-b}-\frac{a}{v^2+2bv-b^2}\frac{d\alpha}{dT}.
    \label{eq:PR_dpdt}
    \end{equation}
    From Eq.~\eqref{eq:PR_alpha},
    \begin{equation}
    \frac{d\alpha}{dT} =-\frac{\kappa\left[1+\kappa\left(1-\sqrt{T_r}\right)\right]}{T_c\sqrt{T_r}}.
    \label{eq:PR_dalpha}
    \end{equation}
    Substituting Eq.~\eqref{eq:PR_dpdt} into
    Eq.~\eqref{eq:app_latent_integral} gives
    \begin{align}
    L={}&RT\int_{v_l}^{v_v}\frac{dv}{v-b}\nonumber\\
    &-aT\frac{d\alpha}{dT}\int_{v_l}^{v_v}\frac{dv}{v^2+2bv-b^2}.
    \label{eq:PR_latent_integrals}
    \end{align}
    The first integral in Eq.~\eqref{eq:PR_latent_integrals} is
    \begin{equation}
    \int_{v_l}^{v_v}\frac{dv}{v-b}=\ln\left(\frac{v_v-b}{v_l-b}\right).
    \label{eq:PR_integral_1}
    \end{equation}
    For the second integral, noting that
    \begin{equation}
    v^2+2bv-b^2 =(v+b)^2-2b^2,
    \end{equation}
    one obtains
    \begin{equation}
    \int\frac{dv}{v^2+2bv-b^2}=\frac{1}{2\sqrt{2}b}\ln\left| \frac{v+b-\sqrt{2}b}{v+b+\sqrt{2}b}\right|.
    \label{eq:PR_integral_indefinite}
    \end{equation}
    Therefore,
    \begin{align}
    \int_{v_l}^{v_v}&\frac{dv}{v^2+2bv-b^2}={}\\
    &\frac{1}{2\sqrt{2}b}\ln\left[ \frac{(v_v+b-\sqrt{2}b)	(v_l+b+\sqrt{2}b)
    }{(v_v+b+\sqrt{2}b)(v_l+b-\sqrt{2}b)}\right].
    \label{eq:PR_integral_2}
    \end{align}
    Substituting Eq.~\eqref{eq:PR_integral_1} and \eqref{eq:PR_integral_2} into Eq.~\eqref{eq:PR_latent_integrals}, the latent heat associated with the PR EOS is finally obtained as
    \begin{align}
    L={}&
    RT\ln\left(\frac{v_v-b}{v_l-b}\right)\nonumber\\
    &-\frac{aT}{2\sqrt{2}b}\frac{d\alpha}{dT}\ln \left[\frac{(v_v+b-\sqrt{2}b)
    (v_l+b+\sqrt{2}b)}{(v_v+b+\sqrt{2}b)(v_l+b-\sqrt{2}b)}\right].
    \label{eq:PR_latent_final}
    \end{align}
    For numerical implementation, it is convenient to express
    Eq.~\eqref{eq:PR_latent_final} directly in terms of the coexistence
    densities. Using $v_l=1/\rho_l$ and $v_v=1/\rho_v$, one obtains
    \begin{align}
    L&={}RT\ln\left[\frac{\rho_l(1-b\rho_v)}{\rho_v(1-b\rho_l)}
    \right]\nonumber\\
    &-\frac{aT}{2\sqrt{2}b}\frac{d\alpha}{dT}\ln\left[\frac{
    \left[1+(1-\sqrt{2})b\rho_v\right]\left[1+(1+\sqrt{2})b\rho_l\right]
    }{\left[1+(1+\sqrt{2})b\rho_v\right]\left[1+(1-\sqrt{2})b\rho_l\right]}
    \right].
    \label{eq:PR_latent_density}
    \end{align}

    Therefore, the latent heat associated with the adopted PR equation of state can be evaluated from the liquid-vapor coexistence properties through the thermodynamic relation derived above. The resulting value, provides a thermodynamic reference for the latent heat implied by the EOS. In the present total enthalpy formulation, however, the latent heat is not restricted to this value and can be specified independently as a caloric property according to the physical problem under consideration~\cite{Zhang2021}.
    
    \subsection{Total enthalpy MRT LB equation}
    The present thermal LB model is formulated directly within the MRT framework. The evolution equation with the MRT collision operator $\hat{\mathbf{M}}$ is given by
    \begin{equation}
        \begin{split}
    &g_\alpha(\bm r+\bm e_\alpha\delta t,t+\delta t)=g_\alpha(\bm r,t) -(\hat{\mathbf{M}}^{-1}\bm S\hat{\mathbf{M}})_{\alpha \beta}(g_\beta(\bm r,t)
    \\&-g_\beta^{eq}(\bm r,t))
    +\delta t\left[\hat{\mathbf{M}}^{-1}\left(\bm I-\frac{\bm S}{2}\right)\hat{\mathbf{M}}\right]_{\alpha\beta}Q_\beta(\bm r,t),
        \end{split}\label{eq:mrt_moment}
    \end{equation}   
    where $\mathbf{S}={diag}(s_0,s_x,s_y,s_E,s_N)$ is the diagonal relaxation matrix. Unlike most previous MRT models that rely on the orthogonal transformation matrix, the present scheme employs a nonorthogonal one, which is simpler and more efficient owing to its sparser structure (more zero elements)~\cite{He2019}. For the D2Q5 lattice, the nonorthogonal transformation matrix $\hat{\mathbf{M}}$ is given by~\cite{Liu2016,LycettBrown2014}
    \begin{equation}
    \hat{\mathbf{M}}=
    \begin{bmatrix}
    1&1&1&1&1\\
    0&1&0&-1&0\\
    0&0&1&0&-1\\
    0&1&1&1&1\\
    0&1&1&1&1
    \end{bmatrix}.\label{eq:M_nonorth}
    \end{equation}
    
    $g_\alpha^{eq}$ is the local equilibrium distribution function given by
    \begin{equation}
    \begin{split}
    g_\alpha^{eq}&=\omega_\alpha\left[(\rho c_p)_{m,\rm ref}( T - T_{\mathrm{ref}})+\frac{\bm e_\alpha\cdot(H\bm u)}{c_s^2}\right]
   \\& +\delta_{\alpha 0}\left(H-(\rho c_p)_{m,\rm ref}( T - T_{\mathrm{ref}})\right),
    \label{eq:distribution function_H}
    \end{split}
    \end{equation}
    where $T_{\rm ref}$ is the reference temperature selected according to the
    specific application. In general, it is chosen as the lower temperature
    within the considered thermal range. Correspondingly, a reference
    volumetric heat capacity, denoted by $(\rho c_p)_{m,\rm ref}$ to ensure dimensional consistency with the volumetric enthalpy $H$. The corresponding weight coefficients for D2Q5 model are respectively given as~\cite{Li2013TLBE}
    \begin{equation}
    \omega_\alpha = 
    \begin{cases} 
    1/3, & \alpha = 0, \\ 
    1/6, & \alpha = 1-4,
    \end{cases}
    \label{eq:d2q5_weights}
    \end{equation}
    the total enthalpy is evaluated by
    \begin{equation}
    H = \sum_{\alpha=0}^{8} g_\alpha. \label{eq:H}
    \end{equation}
    
    Then, the liquid fraction $f_l$ and the temperature $T$ are determined by the total enthalpy, based on the following thermodynamic relations
    \begin{subequations}
    	\begin{align}
    	f_l=\operatorname{clip}\left(\frac{\rho-\rho_v}{\rho_l-\rho_v},0,1\right),\label{eq:f_l}\\
    	T=T_{\rm ref}+\frac{H-f_v\rho_vL}{(\rho c_p)_m}, \label{eq:T}
    	\end{align}
    \end{subequations}
    where $\rho_v$ and $\rho_l$ in Eq.~\eqref{eq:f_l} are the vapor and liquid densities, respectively. In Eq.~\eqref{eq:T}, $f_v = 1 - f_l$ denotes the vapor fraction, $L$ is the latent heat of vaporization, and $(\rho c_p)_m=f_l \rho_l c_{p,l}+f_v \rho_v c_{p,v} $ represents the heat capacity. In Eq.~\eqref{eq:mrt_moment}, $Q_\alpha$ is a correction term defined as
    \begin{equation}
    Q_\alpha=\omega_\alpha \frac{\bm e_\alpha \cdot\partial_t(H\bm u)}{c_s^2}.\label{eq:correction term}
    \end{equation}
    
    In order to recover the macroscopic enthalpy equation, we perform a multiscale analysis of the present model in the following.  Accordingly, the distribution function $g_\alpha$ and the derivatives with respect to time and space are first expanded as~\cite{Guo2013}, 
    \begin{equation}
    g_\alpha=g_\alpha^{(0)}+\varepsilon g_\alpha^{(1)}+\varepsilon^2 g_\alpha^{(2)},
    \partial_t=\varepsilon\partial_{t_1}+\varepsilon^2\partial_{t_2},
    Q_\alpha=\varepsilon Q_\alpha^{(1)},\label{eq:CE}
    \end{equation}
    where $t_1$ is the fast convective scale, $t_2$ is the slow diffusive scale, and is a small parameter which is proportional to the Knudsen number in the classical kinetic theory for fluid ﬂows. And using Taylor expansion of Eq.~\eqref{eq:mrt_moment}, we obtain the following equations at successive orders of $\varepsilon$:
    \begin{subequations}
    	\begin{align}
    	O(\varepsilon^0)&:g_\alpha^{(0)}=g_\alpha^{eq},\label{eq:mrt varepsilon0}\\
    	O(\varepsilon^1)&:
    	\begin{aligned}[t]
    	D_{1 \alpha}g_\alpha^{eq}=&-\frac{1}{ \delta t}(\hat{\mathbf{M}}^{-1}\bm S\hat{\mathbf{M}})_{\alpha \beta}g_\beta^{(1)}\\
    	&+\left [\hat{\mathbf{M}}^{-1}\left(\bm I-\frac{\bm S}{2}\right)\hat{\mathbf{M}}\right]_{\alpha\beta} Q^{(1)}_\beta
    	\end{aligned}\label{eq:mrt varepsilon1}\\
    	O(\varepsilon^2)&:
    	\begin{aligned}[t]
    	\partial_{t_2} g_\alpha^{eq}+D_{1 \alpha}g_\alpha^{(1)}
    	&+\frac{\delta t}{2}D_{1 \alpha}^2g_\alpha^{eq}=\\
    	&-\frac{1}{\delta t}(\hat{\mathbf{M}}^{-1}\bm S\hat{\mathbf{M}})_{\alpha \beta}g_\beta^{(2)},
    	\end{aligned}
  \label{eq:mrt varepsilon2}
    	\end{align}
    \end{subequations}
    where $D_{1\alpha}=\partial_{t1}+d_{1\alpha }=\partial_{t1}+\bm{e_\alpha} \cdot \nabla_1$. Multiplying $\hat{\mathbf{M}}$ on both sides of the above equations, we can obtain the following equations in the moment space:
    \begin{subequations}
    	\begin{align}
    	O(\varepsilon^0)&:\hat{\mathbf{m}}^{(0)}=\hat{\mathbf{m}}^{eq},\label{eq:Mrt varepsilon0}\\
    	O(\varepsilon^1)&:\bm D_{1}\hat{\mathbf{m}}^{eq}=-\frac{1}{ \delta t}{\mathbf{S}}\hat{\mathbf{m}}^{(1)}+
    	\left(\bm I-\frac{1}{2}{\mathbf{S}}\right)\hat{\mathbf{Q}}^{(1)},\label{eq:Mrt varepsilon1}\\
    	O(\varepsilon^2)&:\partial_{t_2} \hat{\mathbf{m}}^{eq}+\bm D_{1}\hat{\mathbf{m}}^{(1)}+\frac{\delta t}{2}\bm D_{1 }^2\hat{\mathbf{m}}^{eq}=-\frac{1}{\delta t}{\mathbf{S}}\hat{\mathbf{m}}^{(2)},\label{eq:Mrt varepsilon2}
    	\end{align}
    \end{subequations}
    where
    \begin{equation}
    \bm{D}_1=\partial_{t_1}+\bm{d}_1,
    \end{equation}
    with
    \begin{equation}
    \begin{aligned}
    \bm{d}_1
    ={}& \hat{\mathbf{M}}
    \operatorname{diag}\bigl[
    \bm{e}_0\cdot\nabla_1,\,
    \bm{e}_1\cdot\nabla_1,\,
    \ldots, \bm{e}_4\cdot\nabla_1
    \bigr]
    \hat{\mathbf{M}}^{-1},
    \end{aligned}
    \end{equation}
    and
    \begin{equation}
    \hat{\mathbf{Q}}^{(1)}
    =
    \left[
    0,\,
    \partial_{t_1}(Hu_x),\,
    \partial_{t_1}(Hu_y),\,
    0,\,
    0
    \right]^{\mathrm T}.
    \end{equation}
    In addition, $\hat{\mathbf{m}}=\hat{\mathbf{M}}\mathbf{g}$ denotes the
    moment vector, and $\hat{\mathbf{m}}^{\mathrm{eq}}$ denotes the
    equilibrium moment vector, defined as
    \begin{equation}
    \bm m^{eq}=\hat{\mathbf{M}}g^{eq}=[H, Hu_x,Hu_y,2c_s^2(\rho c_p)_{m,\rm ref}( T - T_{\mathrm{ref}}),0].\label{eq:meq_d2q5}
    \end{equation}
    In accordance with Eq.~\eqref{eq:Mrt varepsilon1}, the first order equations at the $t_1$ scale may be reformulated. Since only the first three of these equations contribute to the derivation of the macroscopic temperature equation, we present them below and omit the rest.
    \begin{subequations}
    	\begin{gather}
    	\partial_{t_1} H + \partial_{x_1}(H u_x) + \partial_{y_1}(H u_y) = 0, \label{eq:mrt_eps1a} \\
    	\begin{aligned}[t]
    	\partial_{t_1}(H u_x) + c_s^2 (\rho c_p)_{m,\rm ref} &\partial_{x_1} 
    	( T - T_{\mathrm{ref}})
    	= -\frac{s_x}{\delta t} m_x^{(1)}\\
    	&+ \left(1 - \frac{s_x}{2}\right) \partial_{t_1}(H u_x),
    	\end{aligned} \label{eq:mrt_eps1b} \\
    	\begin{aligned}[t]
    	\partial_{t_1}(H u_y) + c_s^2 (\rho c_p)_{m,\rm ref}& \partial_{y_1} ( T - T_{\mathrm{ref}})
    	= -\frac{s_y}{\delta t} m_y^{(1)}
    	\\&+ \left(1 - \frac{s_y}{2}\right) \partial_{t_1}(H u_y).
    	\end{aligned} \label{eq:mrt_eps1c}
    	\end{gather}
    \end{subequations}
    It is worth noting that $\bm{d}_1^2 \hat{\mathbf{m}}^{(0)} = (\mathbf{E} \cdot \nabla_1)(\mathbf{E} \cdot \nabla_1) \hat{\mathbf{m}}^{(0)}$, with $\mathbf{E} = (E_x, E_y)$, where $E_x = \hat{\mathbf{M}} \mathrm{diag}[\bm e_{0x}, \ldots, \bm e_{4x}] \hat{\mathbf{M}}^{-1}$ and $E_y = \hat{\mathbf{M}} \mathrm{diag}[\bm e_{0y}, \ldots, \bm e_{4y}] \hat{\mathbf{M}}^{-1}$. Accordingly, we obtain the following equation for the conservative variable $\theta$ at the $t_2$ scale:
    \begin{equation}
    \begin{aligned}[t]
    &\partial_{t_2}H+\partial_{x_1}m_x^{(1)}+\partial_{y_1}m_y^{(1)}=\\
    &-\frac{\delta t}{2}\partial_{x_1}\left[\partial_{t_1}(Hu_x)
    +c_s^2(\rho c_p)_{m,\rm ref}\partial_{x_1}( T - T_{\mathrm{ref}})\right]\\
    &-\frac{\delta t}{2}\partial_{y_1}\left[\partial_{t_1}(Hu_y)+c_s^2(\rho c_p)_{m,\rm ref}\partial_{y_1}( T - T_{\mathrm{ref}})\right],
    \end{aligned}\label{eq:mrt_eps2}
    \end{equation}
    Substituting Eq.~\eqref{eq:mrt_eps1b} and Eq.~\eqref{eq:mrt_eps1c} into Eq.~\eqref{eq:mrt_eps2}, the time derivative terms cancel out, yielding
    \begin{equation}
    \begin{aligned}[t]
    \partial_{t_2}H&={}
    \partial_{x_1}\left[c_s^2(\rho c_p)_{m,\rm ref}\delta t\left(\frac1{s_x}-\frac12\right)\partial_{x_1}T\right]
    \\&+\partial_{y_1}\left[c_s^2(\rho c_p)_{m,\rm ref}\delta t\left(\frac1{s_y}-\frac12\right)\partial_{y_1}T\right],
    \end{aligned}\label{eq:mrt_eps2_H}
    \end{equation}
    Based on Eq.~\eqref{eq:mrt_eps1a} and Eq.~\eqref{eq:mrt_eps2_H}, we can get
    \begin{equation}
    \partial_t H+\nabla\cdot(H\bm u)=\nabla \cdot (\lambda \nabla T),\label{eq:macroscopic_temperature}
    \end{equation}
    where $\lambda=c_s^2(\rho c_p)_{m,\rm ref}\delta t\left(\frac1{s_x}-\frac12\right)=c_s^2(\rho c_p)_{m,\rm ref}\delta t$\\$\left(\frac1{s_y}-\frac12\right)$.

    The foregoing multiscale analysis demonstrates that the present MRT formulation recovers the target macroscopic total enthalpy equation without spurious deviation terms. The thermal diffusion term is recovered through the lattice Boltzmann evolution, while the nonlinear thermodynamic source term required in conventional temperature-based formulations is avoided. These features preserve the locality and computational efficiency of the LB scheme and provide a consistent framework for liquid-vapor phase change heat transfer.

	\section{PSEUDOPOTENTIAL LB MODEL}
	\label{sec:PSEUDOPOTENTIAL LB MODEL}
	Among the various LB multiphase approaches, the Shan and Chen pseudopotential model~\cite{Shan1993} stands out for its intuitive physical picture and straightforward coupling with non-ideal gas equations of state. Nevertheless, achieving satisfactory thermodynamic consistency has motivated substantial modifications to the original scheme over the past two decades ~\cite{Li2013,Huang2016ThirdOrder,Kupershtokh2009}. For the purposes of this study, we adopt the improved D2Q9 MRT pseudopotential model due to Li et al.~\cite{Li2013}, and the corresponding evolution equation for the density population $f_\alpha$ reads:
	\begin{gather}
	\begin{aligned}[t]
	f_\alpha(\mathbf{r}+\mathbf{e_\alpha} \delta t,t+\delta t)-&f_\alpha(\mathbf{r},t)=-\bar{\Lambda}_{\alpha \beta}(f_\beta(\mathbf{r},t)-f_\beta^{eq}(\mathbf{r},t))
	\\
	&+\delta t(\mathbf{S_\alpha}(\mathbf{r},t)-\frac{1}{2} \bar{\Lambda}_{\alpha \beta}\mathbf{S_\beta}(\mathbf{r},t)), \label{eq:lb_evolution}
	\end{aligned}
	\end{gather}
	in which $\mathbf{e_\alpha}$ is the discrete velocity at position $\mathbf{r}$ and time $t$. For the D2Q9 model considered here, the discrete velocities are defined as~\cite{Kruger2017}
	\begin{gather}
	\mathbf{e}_\alpha =
	\begin{cases} 
	(0, 0), & \alpha = 0, \\
	\left(\cos\left[\frac{(i-1)\pi}{2}\right], \sin\left[\frac{(i-1)\pi}{2}\right]\right) c, & \alpha=1-4, \\
	\left(\cos\left[\frac{(2i-9)\pi}{4}\right], \sin\left[\frac{(2i-9)\pi}{4}\right]\right) \sqrt{2}\,c, & \alpha = 5-8,
	\end{cases}
	\label{eq:d2q9_velocity}
	\end{gather}
	where $c=dx/dt$ is the lattice speed with $x$ and $t$ denoting the lattice spacing and time step. $f_\alpha^{eq}(\mathbf{r},t)$ is the equilibrium distribution given by~\cite{Chen1998}
	\begin{gather}
	f_\alpha^{eq}(\mathbf{r},t) = \omega_\alpha \rho \left[	1 + \frac{\mathbf{e}_\alpha \cdot \mathbf{u}}{c_s^2}	+ \frac{(\mathbf{e}_\alpha \cdot \mathbf{u})^2}{2c_s^4}	- \frac{|\mathbf{u}|^2}{2c_s^2}\right],
	\label{eq:equilibrium}
	\end{gather}
	where $\omega_\alpha$ is the weight coefficient given by $\omega_0 = 4/9,\omega_{1-4} = 1/9,\omega_{5-8} = 1/36,u = (u_x,u_y)$ is the velocity, and $c_s^2 = c^2/3$ is the sound speed. The collision matrix $\boldsymbol{\Lambda}$ is transformed into the moment space by $\bar{\boldsymbol{\Lambda}} = \mathbf{M}^{-1} \boldsymbol{\Lambda} \mathbf{M}$, where $\mathbf{M}$ is the transformation matrix given by~\cite{Lallemand2000}
	\begin{gather}
	\mathbf{M}=
	\begin{bmatrix}
	1 & 1 & 1 & 1 & 1 & 1 & 1 & 1 & 1\\
	-4 & -1 & -1 & -1 & -1 & 2 & 2 & 2 & 2\\
	4 & -2 & -2 & -2 & -2 & 1 & 1 & 1 & 1\\
	0 & 1 & 0 & -1 & 0 & 1 & -1 & -1 & 1\\
	0 & -2 & 0 & 2 & 0 & 1 & -1 & -1 & 1\\
	0 & 0 & 1 & 0 & -1 & 1 & 1 & -1 & -1\\	
	0 & 0 & -2 & 0 & 2 & 1 & 1 & -1 & -1\\	
	0 & 1 & -1 & 1 & 1 & 0 & 0 & 0 & 0\\	
	0 & 0 & 0 & 0 & 0 & 1 & -1 & 1 & -1		
	\end{bmatrix}
		\label{eq:transformation matrix}.
	\end{gather}
	In addition, when Eq.~\eqref{eq:lb_evolution} is multiplied by the transformation matrix $\mathbf{M}$, the collision step originally implemented in velocity space is consequently performed in moment space,
	\begin{gather}
		\begin{aligned}[t]
	\mathbf{m}^*(\mathbf{r},t)=\mathbf{m}(\mathbf{r},t)-&\Lambda(\mathbf{m}(\mathbf{r},t)-\mathbf{m}^{eq}(\mathbf{r},t))
	\\
	&+\delta t(\mathbf{I}-\frac{\Lambda}{2})\bar{\mathbf{S}}(\mathbf{r},t),
		\end{aligned}
		\label{eq:lb_evolution_M}
	\end{gather}
	while the streaming step is still implemented in the velocity space,
	\begin{gather}
	f_\alpha(\mathbf{r}+\mathbf{e_\alpha} \delta t,t+\delta t)=	f^*_\alpha(\mathbf{r},t),
		\label{eq:lb_collision}
	\end{gather}
	where $\mathbf{m}(\mathbf{r},t)=\mathbf{M}\mathbf{f}$ is the rescaled moment, with $\mathbf{f}(\mathbf{r},t)$\\
	$=[f_0(\mathbf{r},t), f_1(\mathbf{r},t), \ldots, f_8(\mathbf{r},t)]^T$, and $f^*_\alpha(\mathbf{r},t)=\mathbf{M}^{-1}\mathbf{m}^*$	\\
    $(\mathbf{r},t)$ denotes the postcollision distribution function. He\\
	re, $\boldsymbol{\Lambda}=\mathrm{diag}(1/\tau_c, 1/\tau_e, 1/\tau_\xi, 1/\tau_c, 1/\tau_q, 1/\tau_c, 1/\tau_q, 1/\tau_\nu,$\\$ 1/\tau_\nu)$ is the diagonal relaxation matrix, and $\mathbf{m}^{\mathrm{eq}}$ represents the equilibrium distribution function in moment space, following Lallemand and Luo~\cite{Lallemand2000}, the equilibrium moments are given by
	\begin{gather}
	\begin{aligned}[t]
	\mathbf{m}^{eq}(\mathbf{r}, t) = \rho& [1, \,-2 + 3|\mathbf{u}|^2, \,1 - 3|\mathbf{u}|^2, \,	u_x, \,-u_x, \,	u_y,\\ 
	&\,	-u_y, \,	u_x^2 - u_y^2, \,	u_x u_y]^{\mathrm{T}}.
	\end{aligned}
	\label{eq:meq}
	\end{gather}
	In addition, $\bar{\mathbf{S}}$ is the forcing term in the moment space, which satisfies $\bar{\mathbf{S}}=\mathbf{M}\mathbf{S}$. According to Li et al.~\cite{Li2013}, it can be written as
	\begin{gather}
	\bar{\mathbf{S}} = 
	\begin{bmatrix}
	0 \\
	6\mathbf{u} \cdot \mathbf{F} + \dfrac{12 \sigma |\mathbf{F}_{ads}|^2}{\psi^2 \delta t \left(\tau_e - 0.5\right)} \\
	-6\mathbf{u} \cdot \mathbf{F} - \dfrac{12 \sigma |\mathbf{F}_{ads}|^2}{\psi^2 \delta t \left(\tau_\xi - 0.5\right)}\\
	F_x\\
	-F_x\\
	F_y\\
	-F_y\\
	2(u_x F_{x} - u_y F_{y})\\
	u_x F_{y} + u_y F_{x}
	\end{bmatrix},
	\label{eq:F_moment}
	\end{gather}
	here $\sigma$ is an adjustable parameter employed to enforce thermodynamic consistency and $\mathbf{F}=(F_x,F_y)$ is the total force, and it is defined as $\mathbf{F}=\mathbf{F}_{ads}+\mathbf{F}_b$, where $\mathbf{F}_b$ is the external force and $\mathbf{F}_{ads}$ is the intermolecular interaction force. The macroscopic and mesoscopic parameters are related by
	\begin{subequations}
		\begin{align}
		\rho &= \sum_{\alpha=0}^{8} f_\alpha, \label{eq:rho} \\
		\rho \mathbf{u} &= \sum_{i=0}^{8} \mathbf{e}_\alpha f_\alpha + \frac{\delta t}{2} \mathbf{F}. \label{eq:momentum}
		\end{align}
	\end{subequations}

	In the pseudopotential lattice Boltzmann model, both the nonideal equation of state and the surface tension are captured through the interaction force $\mathbf{F}_{ads}$. With the nearest-neighbor interactions on the D2Q9 lattice, this force is given by~\cite{Li2014}
	\begin{gather}
	\mathbf{F}_{ads}(\mathbf{r}, t) = -G \psi(\mathbf{r}, t) \sum_{\alpha=1}^{8} w(|\mathbf{e}_{\alpha}|^2) \, \psi(\mathbf{r} + \mathbf{e}_{\alpha}, t) \, \mathbf{e}_{\alpha},
	\label{eq:interaction_force_general}
	\end{gather}
	where $\psi(\mathbf{r})$ is the interaction potential, $G$ is the interaction strength whose value is usually set to be $-1$ in many simulations, and $w(|\mathbf{e}_{\alpha}|^2)$ are the weights. For this case, the weights $w(|\mathbf{e}_{\alpha}|^2)$ are $w (1) = 1/3 $ and $w (2) = 1/12$. And $\psi(\mathbf{r})$ is the pseudopotential function, given by $\psi(\rho) = \sqrt{2(p_{\mathrm{EOS}} - \rho c_s^2) / (G c^2)}$, with $p_{\mathrm{EOS}}$ being the nonideal equation of state~\cite{Yuan2006}. In the present work, the PR EOS is used, which is given as
	\begin{equation}
	p_{\mathrm{EOS}}=\frac{\rho R T}{1-b\rho}-\frac{a\alpha(T)\rho^2}{1+2b\rho-b^2\rho^2},
	\label{eq:PR_EOS}
	\end{equation}
	where $R$ is the gas constant,
	$\alpha(T)=[1+(0.37464+
	1.54226\omega-0.26992\omega^2)(1-\sqrt{T/T_c})]^2,a=0.45724R^2T_c^2/$\\
	$p_c,b=0.07780RT_c/p_c$, and $T_c$, $p_c$, and $\omega$ denote the critical temperature, critical pressure, and acentric factor, respectively. We utilize $a = 1/49, b = 2/21$, and $R = 1$ in the present study and choose $\omega = 0.344$. It should be noted that the approximation adopted in the thermal formulation does not imply neglect of the EOS pressure in the pseudopotential model, where it remains essential for describing liquid-vapor phase separation and interfacial mechanical equilibrium.
			
\begin{comment}	
     In the present work, the CS EOS is used, which is given as
    \begin{gather}
	{p}_{\mathrm{{EOS}}} = {\rho RT}\frac{1 + {b\rho }/4 + {\left( b\rho /4\right) }^{2} - {\left( b\rho /4\right) }^{3}}{{\left( 1 - b\rho /4\right) }^{3}}- a{\rho }^{2},	\label{eq:CS_EOS}
    \end{gather}
    where $R$ is the gas constant, $a = {0.4963}{R}^{2}{T}_{c}^{2}/{p}_{c}$, and $b = {0.18727R}{T}_{c}/{p}_{c}$, with $T_c$ and $p_c$ representing the critical temperature and the critical pressure, respectively. we choose $a = 1/2, b = 4$, and $R = 1$, which have been widely used in previous works.
\end{comment}    
    
    \section{RESULTS AND DISCUSSION}
    	\label{sec:RESULTS AND DISCUSSION}
    In this section, we select several benchmark liquid-vapor phase change problems to validate the performance of the proposed LB model. These typical examples include the $D^2$ law of droplet evaporation, droplet evaporation on a heated surface, and bubble nucleation and departure in nucleate boiling. Unless otherwise stated, the non-hydrodynamic relaxation parameters for the velocity field are chosen following Li et al.~\cite{Li2013}, while the relaxation rates associated with the shear modes are determined by the prescribed kinematic viscosity. Accordingly,
    $\boldsymbol{\Lambda}=\mathrm{diag}(1.0, 1.1, 1.1, 1.0, 1.1, 1.0, 1.1, 1/0.74, 1/0.74)$, and those for the enthalpy field are set to $\mathbf{S}={diag}(0.0,s_x,s_y,1.0,1.0)$ in all simulations, where $s_x,s_y$ are determined by the thermal diffusivity $\lambda=c_s^2(\rho c_p)_{m,\rm ref}\delta t\left(\frac1{s_x}-\frac12\right)=c_s^2(\rho c_p)_{m,\rm ref}\delta t\left(\frac1{s_y}-\frac12\right)$, with the reference volumetric heat capacity given by $(\rho c_p)_{m,\rm ref}=\rho_v c_{p,v}$.
    
    \subsection{Validation of the $D^2$ law}
    To assess the quantitative accuracy of the proposed LB model for
    liquid-vapor phase change, we consider the benchmark problem of droplet
    evaporation in open space, which has been widely employed to evaluate
    interfacial heat and mass transfer in liquid-vapor phase change models~\cite{Qiao2026}. In this classic configuration, the evolution of the droplet diameter is governed by the $D^2$ law, which predicts a linear
    decrease of $D^2$ with time~\cite{Zheng2018}.
    \begin{equation}
   (\frac{D}{D_0})^2 =1 - K t,\label{eq:D2}
    \end{equation}
    where $D_0$ is the initial droplet diameter, $t$ is the evaporation time, and $K$ is the slope indicating the evaporation rate.
    
    The simulation is conducted in a $200 \times 200$ square computational domain, at the center of which a droplet with an initial diameter of $D_0 = 60$ is positioned. The droplet is initially maintained at the saturation temperature $T_{{sat}}=0.86 T_c$, and the ambient region is occupied by superheated vapor at a uniform temperature $T_g=1.0T_c$. The density are set according to
    \begin{equation}
    \rho=\frac{(\rho_l + \rho_g)}{2} -\frac{(\rho_l - \rho_g)}{2} tanh(\frac{2.0 * (r - D_0 / 2)}{W} ),\label{eq:diffuse}
    \end{equation}
    where $r=\sqrt{(x-x_0)^2+(y-y_0)^2}$ is the distance to the droplet center, $(x_0,y_0)=(100,100)$ and $W=4.0$. In this case, the droplet is expected to evaporate due to the temperature gradient around the liquid-vapor interface. For the velocity field, the periodic boundary condition is employed to determine the unknown density distribution functions outside the computational domain~\cite{Kruger2017}. For the temperature field, a Dirichlet boundary condition (i.e., $T = T_g$) is imposed, which can be readily realized by the nonequilibrium extrapolation scheme~\cite{Guo2002}. In the whole computational domain, the thermal conductivity, the specific heat at constant volume, and the latent heat of vaporization are set to \(\lambda=0.1\) and $c_{p,v}=c_{p,l} = 1.0$. This choice yields a thermal relaxation time sufficiently away from its lower stability limit and provides a stable thermal transport regime for the present simulations. Furthermore, the latent heat associated with the adopted PR EOS is evaluated from Eq.~\eqref{eq:PR_latent_density}, yielding $L=0.1906$, which agrees with the value computed from the Gong and Cheng's~\cite{Gong2013} formula.
    
\begin{figure}[htbp]
	\centering
	% 第一行
	%	\hspace*{0.08\textwidth}  % 左侧留白
	\begin{minipage}[b]{0.15\textwidth}
		\centering
		\includegraphics[trim=5cm 5cm 6cm 5cm, clip=true, width=0.95\textwidth]{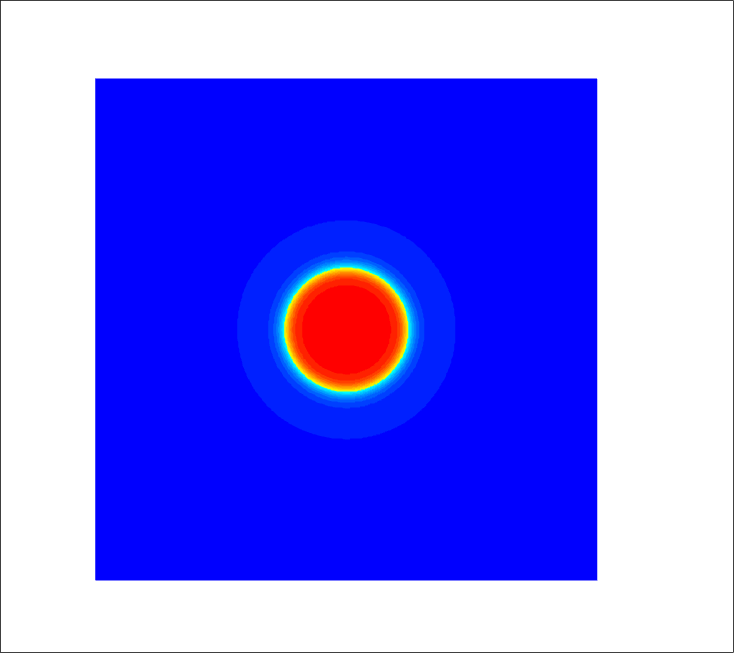}
	\end{minipage}
	%	\hspace*{0.002\textwidth}
	\begin{minipage}[b]{0.15\textwidth}
		\centering
		\includegraphics[trim=5cm 5cm 6cm 5cm, clip=true, width=0.95\textwidth]{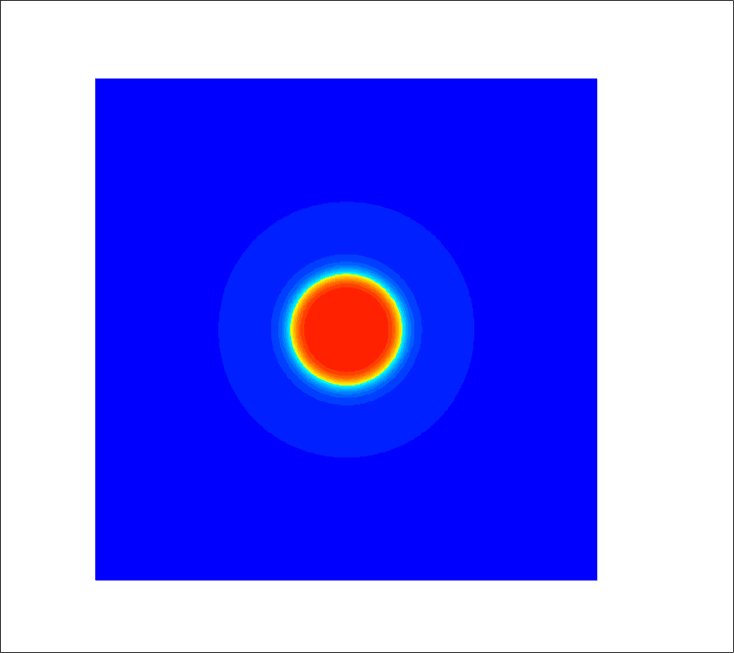}
	\end{minipage}
	% 	\hspace*{0.002\textwidth}
	\begin{minipage}[b]{0.15\textwidth}
		\centering
		\includegraphics[trim=5cm 5cm 6cm 5cm, clip=true, width=0.95\textwidth]{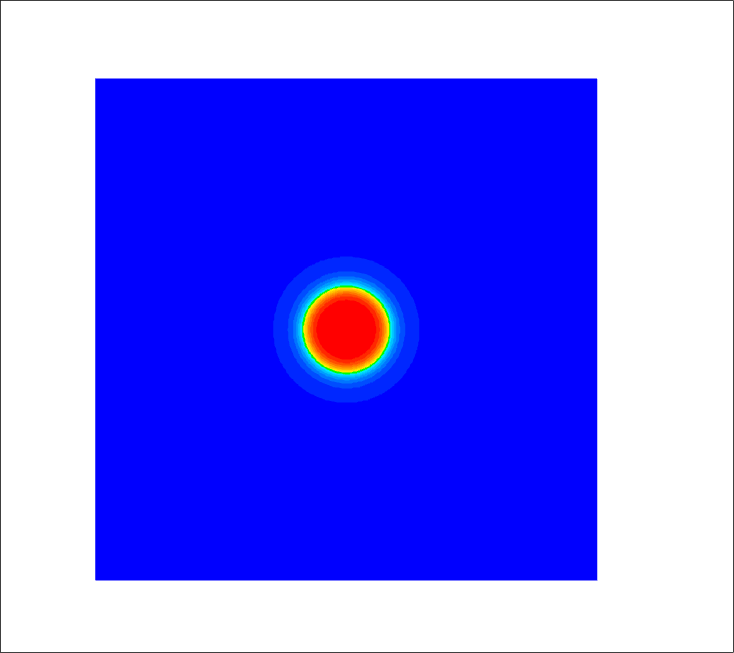}
	\end{minipage}
	% \hspace*{0.08\textwidth}  % 右侧留白
	\vspace{0.005\textheight} 
	\textbf{(a)}
	
	% 第二行
	% 	\hspace*{0.08\textwidth}  % 左侧留白
	\begin{minipage}[b]{0.15\textwidth}
		\centering
		\includegraphics[trim=5cm 5cm 6cm 5cm, clip=true, width=0.95\textwidth]{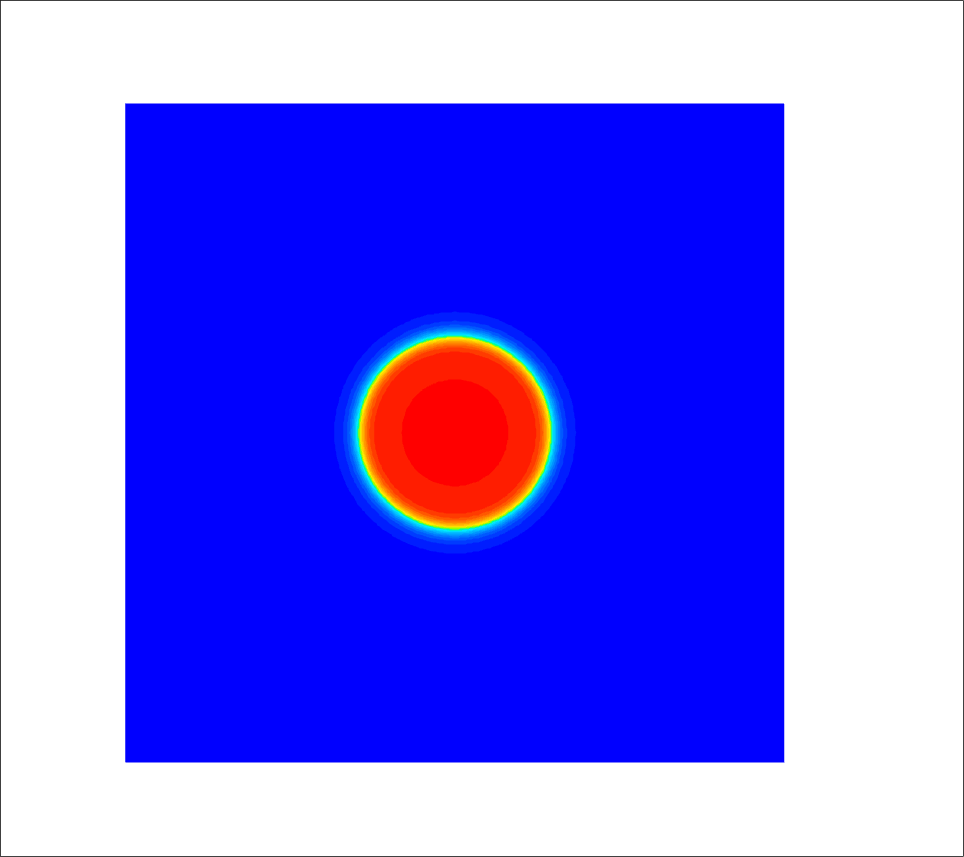}
	\end{minipage}
	%	\hspace*{0.002\textwidth}
	\begin{minipage}[b]{0.15\textwidth}
		\centering
		\includegraphics[trim=5cm 5cm 6cm 5cm, clip=true, width=0.95\textwidth]{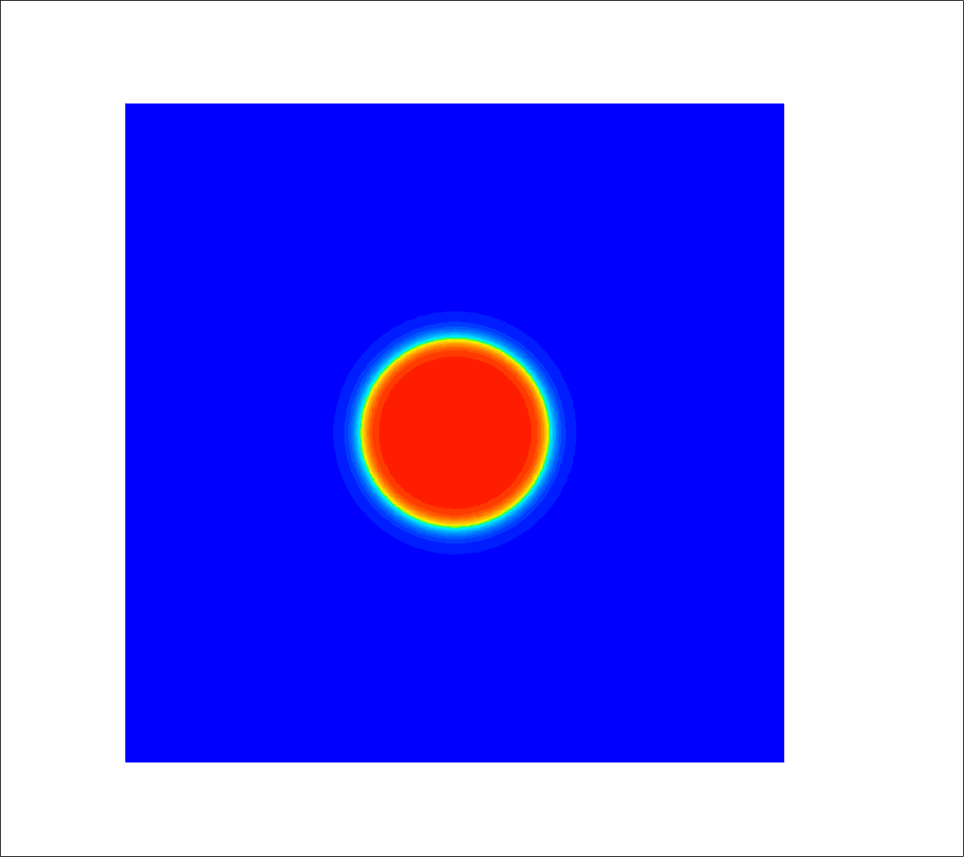}
	\end{minipage}
	% 	\hspace*{0.002\textwidth}
	\begin{minipage}[b]{0.15\textwidth}
		\centering
		\includegraphics[trim=5cm 5cm 6cm 5cm, clip=true, width=0.95\textwidth]{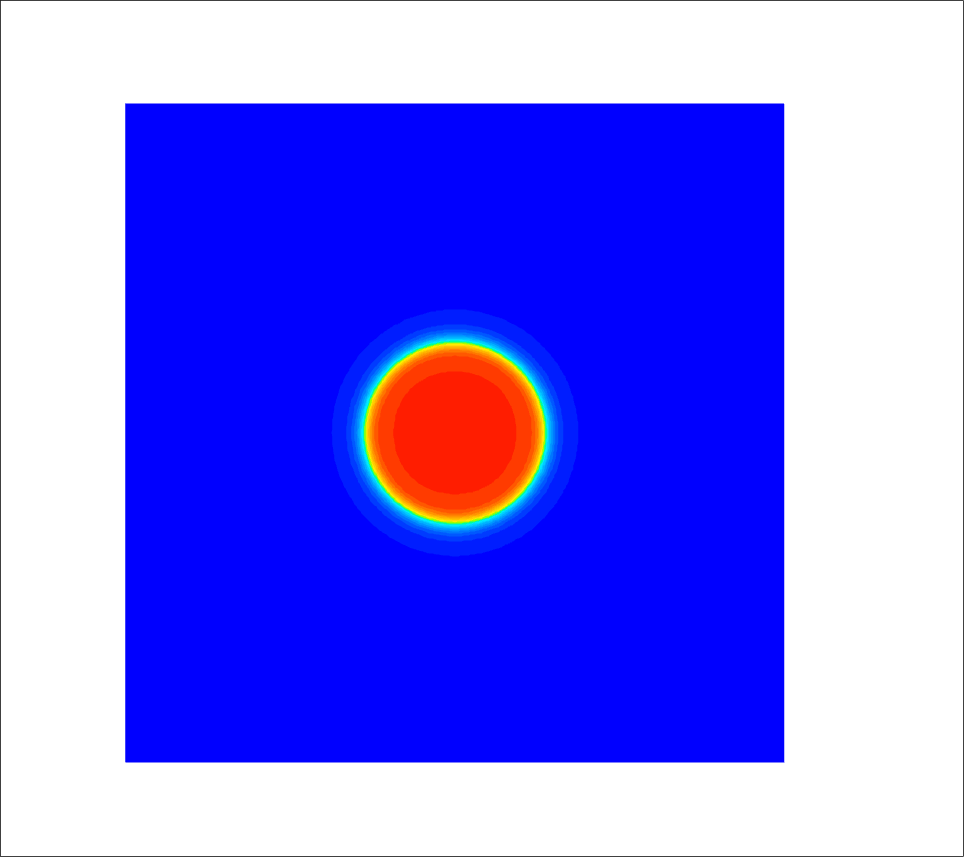}
	\end{minipage}
	%  	\hspace*{0.08\textwidth}  % 右侧留白
	\vspace{0.005\textheight} 
	\textbf{(b)}
	
	% 第三行
	%  	\hspace*{0.08\textwidth}  % 左侧留白
	\begin{minipage}[b]{0.15\textwidth}
		\centering
		\includegraphics[trim=5cm 5cm 6cm 5cm, clip=true, width=0.95\textwidth]{D2_t=0.1.png}
	\end{minipage}
	% 	\hspace*{0.002\textwidth}
	\begin{minipage}[b]{0.15\textwidth}
		\centering
		\includegraphics[trim=5cm 5cm 6cm 5cm, clip=true, width=0.95\textwidth]{D2_t=0.25.png}
	\end{minipage}
	%  	\hspace*{0.002\textwidth}
	\begin{minipage}[b]{0.15\textwidth}
		\centering
		\includegraphics[trim=5cm 5cm 6cm 5cm, clip=true, width=0.95\textwidth]{D2_t=0.5.png}
	\end{minipage}
	%  	\hspace*{0.08\textwidth}  % 右侧留白
	\vspace{0.005\textheight} 
	\textbf{(c)}
	\caption{Comparisons of the density contours given by the Gong-Cheng model (a), the Li et al.’s model (b), and the present LB model (c) at $t^*=0.1,t^*=0.25$ ,$t^*=0.5$ (from left to right).}
	\label{fig:D2_duibi}
\end{figure}

\begin{figure}[htbp]
	\centering
	\includegraphics[width=0.4\textwidth]{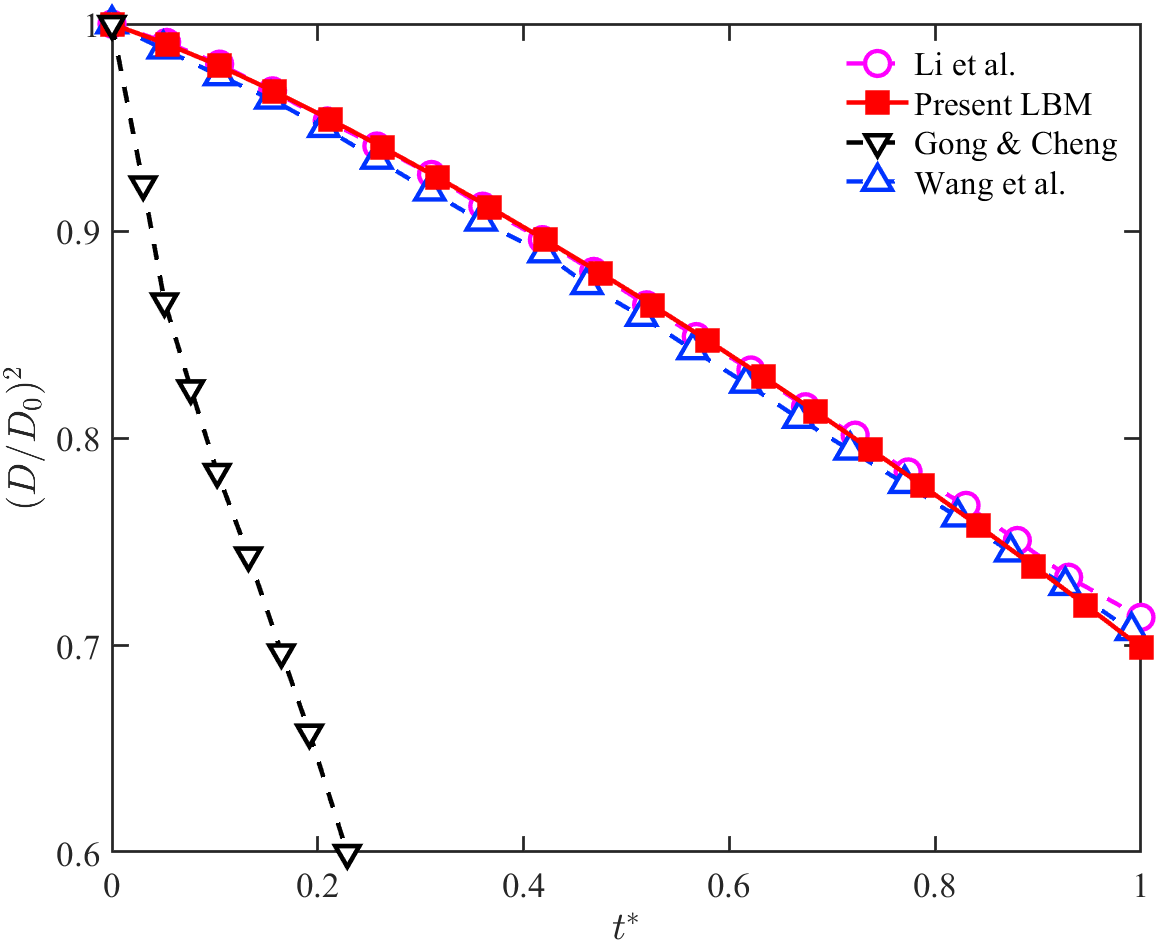}
	\caption{Variation of the square of the dimensionless diameter $(D/D_0)^2$ with the dimensionless time $t^*$.}
	\label{fig:D2_duibi_line}
\end{figure}

    Figure~\ref{fig:D2_duibi} presents the density contours at dimensionless time $t^* = t/t_{total}$, the results obtained from the Gong-Cheng model~\cite{Gong2012} and the Li et al.’s model~\cite{Li2017} are also included in this ﬁgure. It is clear that the simulation results using the present LBM agree well with the Li et al.’s model. Meanwhile the variation of the square of the dimensionless diameter $(\frac{D}{D_0})^2$ with the dimensionless time $t^*$ is displayed in Figure~\ref{fig:D2_duibi_line}. This figures show that the present LB model yields an evaporation process similar to that of Li et al.~\cite{Li2017} and Wang et al.~\cite{Wang2022}, indicating its good accuracy. Meanwhile, the evaporation rate predicted by Gong and Cheng's model~\cite{Gong2012} is much higher than that of the other two LB models. These comparisons confirm the accuracy of the present LB model in predicting droplet evaporation.

\begin{figure}[htbp]
   	\centering
   	\includegraphics[width=0.4\textwidth]{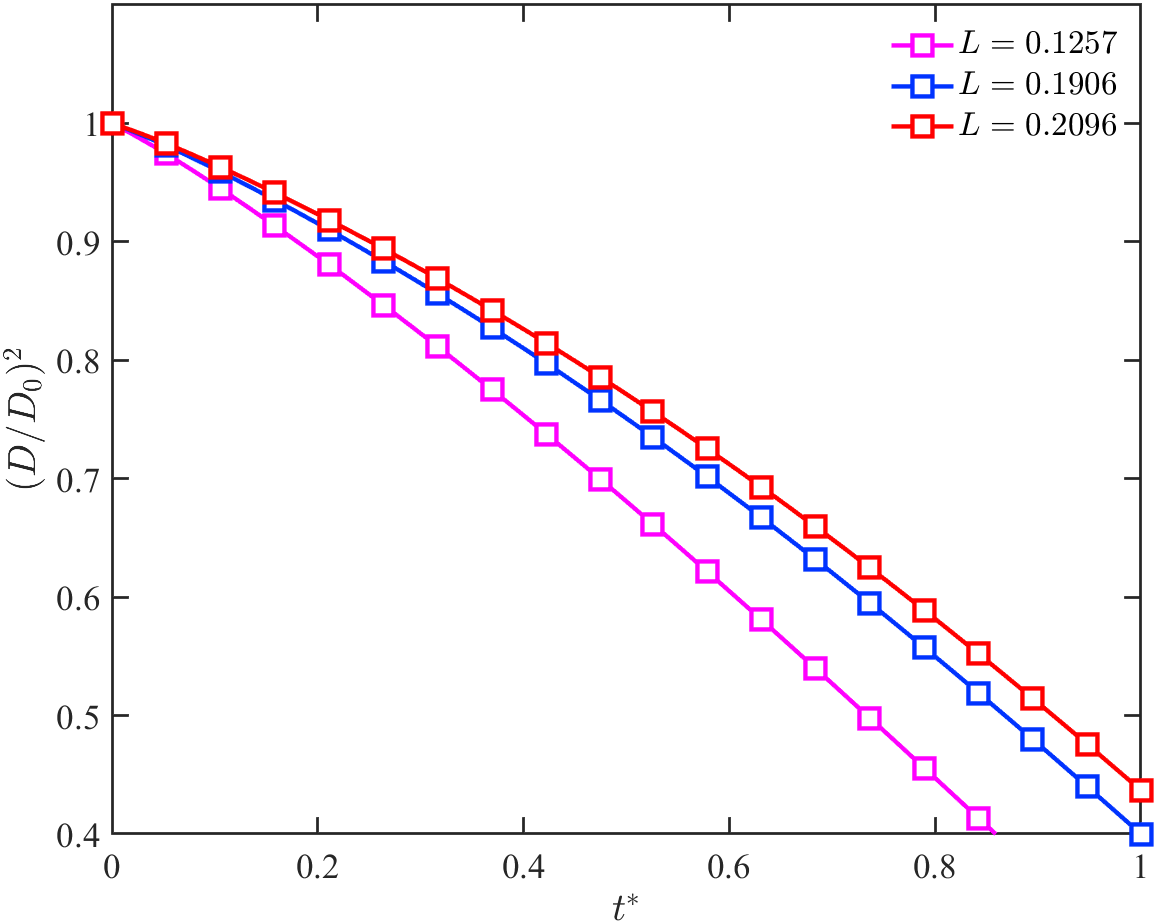}
   	\caption{Variation of the square of the dimensionless diameter $(D/D_0)^2$ with the dimensionless time $t^*$ for different latent heats L.}
   	\label{fig:D2_Lduibi}
   \end{figure}

    To further examine the capability of the present LB model to independently specify the latent heat and to investigate its influence on droplet evaporation, three different latent heats, $L=0.1257$, $0.1906$, and $0.2096$, are considered. The corresponding evaporation behaviors are compared in Fig.~\ref{fig:D2_Lduibi}. The dimensionless time in the figure is defined as 
   $t^*=t/t_{total}$, where $t_{total}$ is the total evaporation time for $L=0.1906$. It can be seen that $(D/D_0)^2$ decreases approximately linearly with time for all the considered cases, indicating that the characteristic $D^2$ evaporation behavior is well preserved. Meanwhile, the evaporation rate decreases noticeably with increasing latent heat. The droplet evaporates fastest for $L=0.1257$, whereas the slowest evaporation is observed for $L=0.2096$. This behavior is physically reasonable since a larger latent heat requires more energy for the vaporization of a given amount of liquid. These results further demonstrate that the present total enthalpy LB model can flexibly specify the latent heat.
   
   \subsection{Droplet evaporation on heated surface}
   Since solid surfaces are commonly encountered in liquid-vapor phase transition problems, we next consider droplet evaporation on a heated surface, which serves as a canonical benchmark for validating the present LB model for liquid-vapor phase change. We consider a $200 \times 100$ computational domain, in which a droplet of initial radius $R_0=35$ and temperature $T_{sat}=0.86T_c$ is positioned on a heated surface, surrounded by saturated vapor.The bottom wall is treated as a solid wall with an equilibrium contact angle of $120^{\circ}$ and is maintained at a temperature of $T_h=0.87T_c$. For the hydrodynamic field, the no-slip condition at the bottom wall is implemented using the bounce-back scheme, while the non-equilibrium extrapolation scheme is employed for the total-enthalpy distribution functions. The open boundary condition is imposed at the top boundary, whereas periodic boundary conditions are applied in the horizontal direction. To describe the fluid-solid interactions, we employ the geometric formulation proposed by Ding and Spelt~\cite{Ding2007} in the subsequent simulations, which is given by 
   \begin{equation}
   \rho_{x,0} = \rho_{x,2} + \tan\left(\frac{\pi}{2} - \theta^{eq}\right)\left|\rho_{x+1,1} - \rho_{x-1,1}\right|,
   \label{eq:density at the wall}
   \end{equation}
   where $\theta^{eq}$ is the equilibrium contact angle, and $\rho_{x,0}$ is the density at the ghost layer near the wall. The subscripts $x$ and $y$ denote the tangential and normal directions to the wall, respectively. The latent heat of vaporization is $L=0.1906$,% the thermal conductivity is \(\lambda=1.6\) 
   and the specific heat at constant volume is calculated according $c_{p,l}$ to $Ja=c_{p,l}(T_h-T_{sat})/L=0.036$~\cite{Zhang2021}. The simulation is first run without evaporation until the droplet contact angle reaches the prescribed value.
   
\begin{figure}[htbp]
   	\centering
   	\begin{minipage}{0.5\textwidth}  % 左右各留4%，图整体更大
   		\centering
   		\begin{subfigure}[b]{0.3\textwidth}   % 子图宽度增大
   			\centering
   			\includegraphics[trim=5cm 3cm 6cm 13cm, clip=true, width=\textwidth]{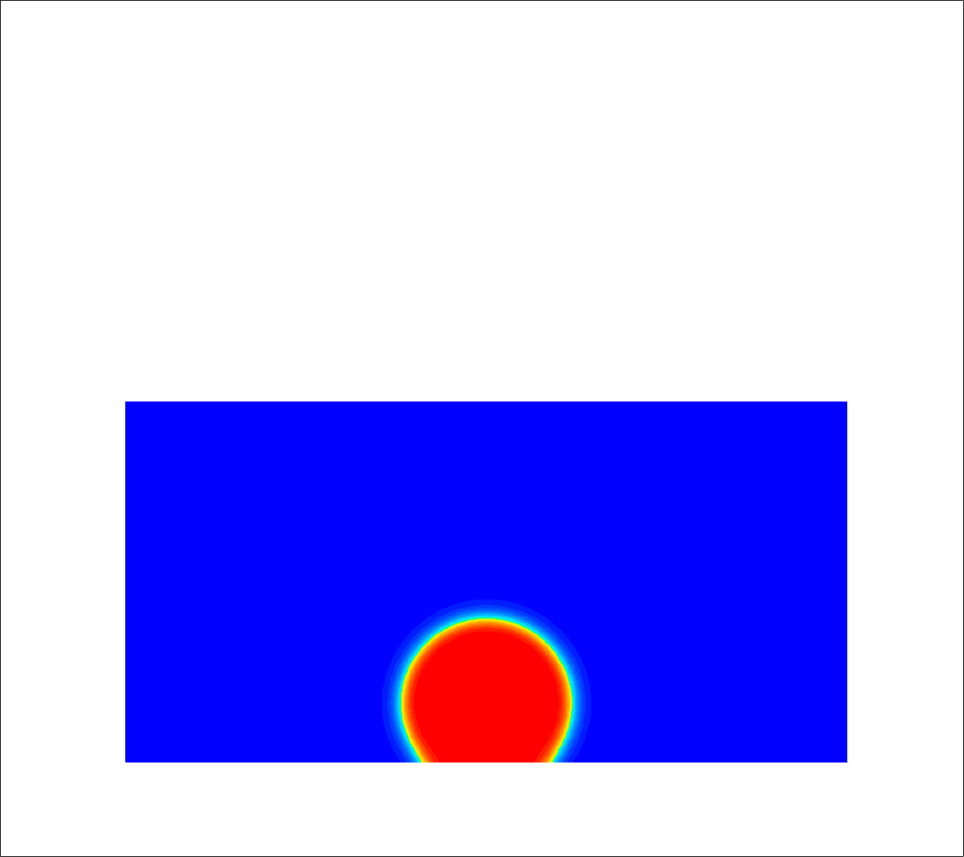}
   			\caption*{(a)}
   			\label{fig:heated surface_sub1}
   		\end{subfigure}%
   		\hspace{0.01\textwidth}%              % 间距极小（0.5%）
   		\begin{subfigure}[b]{0.3\textwidth}
   			\centering
   			\includegraphics[trim=5cm 3cm 6cm 13cm, clip=true, width=\textwidth]{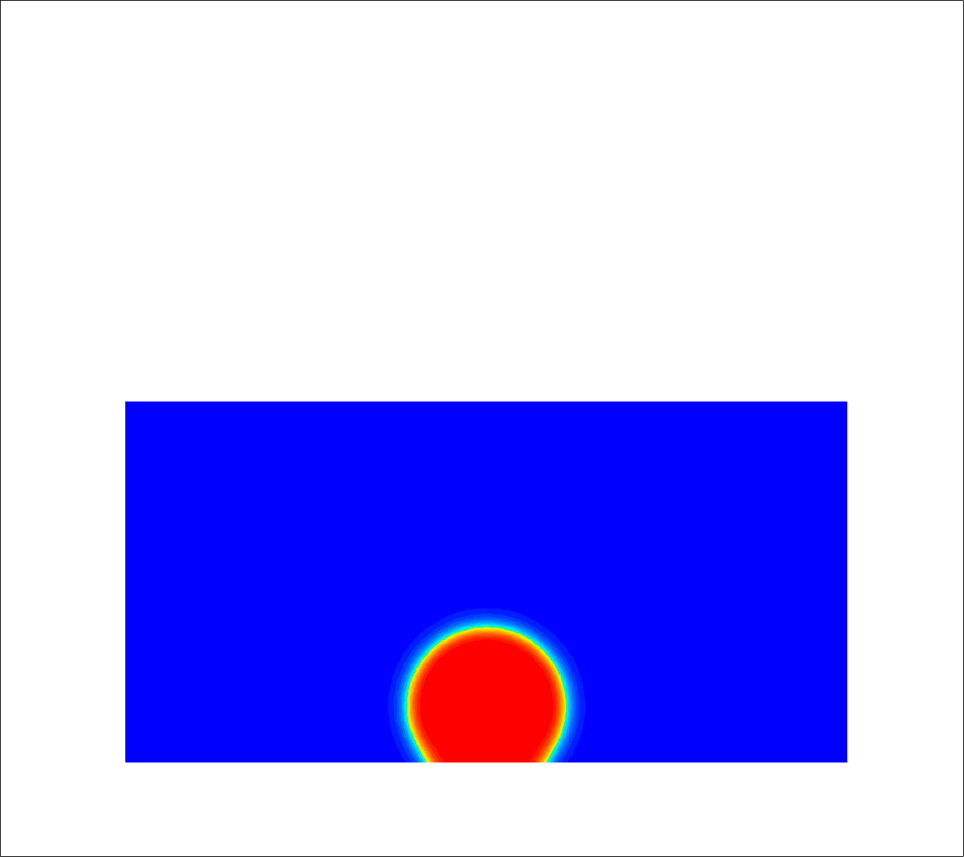}
   			\caption*{(b)}
   			\label{fig:heated surface_sub2}
   		\end{subfigure}%
   		\hspace{0.01\textwidth}%              % 间距极小（0.5%）
   		\begin{subfigure}[b]{0.3\textwidth}
   			\centering
   			\includegraphics[trim=5cm 3cm 6cm 13cm, clip=true, width=\textwidth]{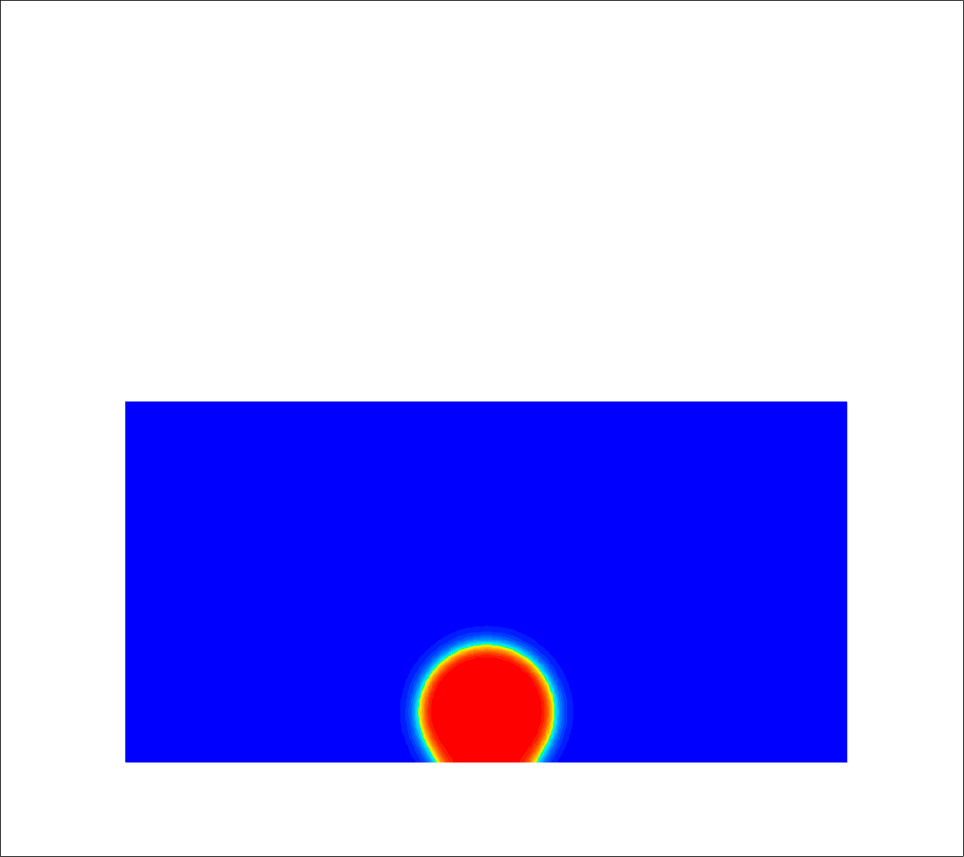}
   			\caption*{(c)}
   			\label{fig:heated surface_sub3}
   		\end{subfigure}
   	\end{minipage}
   	\caption{Droplet evaporation on a solid surface at $t^* =0.1$ (a), $t^* = 0.25$ (b), and $t^* =0.5$ (c).}
   	\label{fig:heated surface}
   \end{figure}

\begin{figure}[htbp]
   	\centering
   	\includegraphics[width=0.4\textwidth]{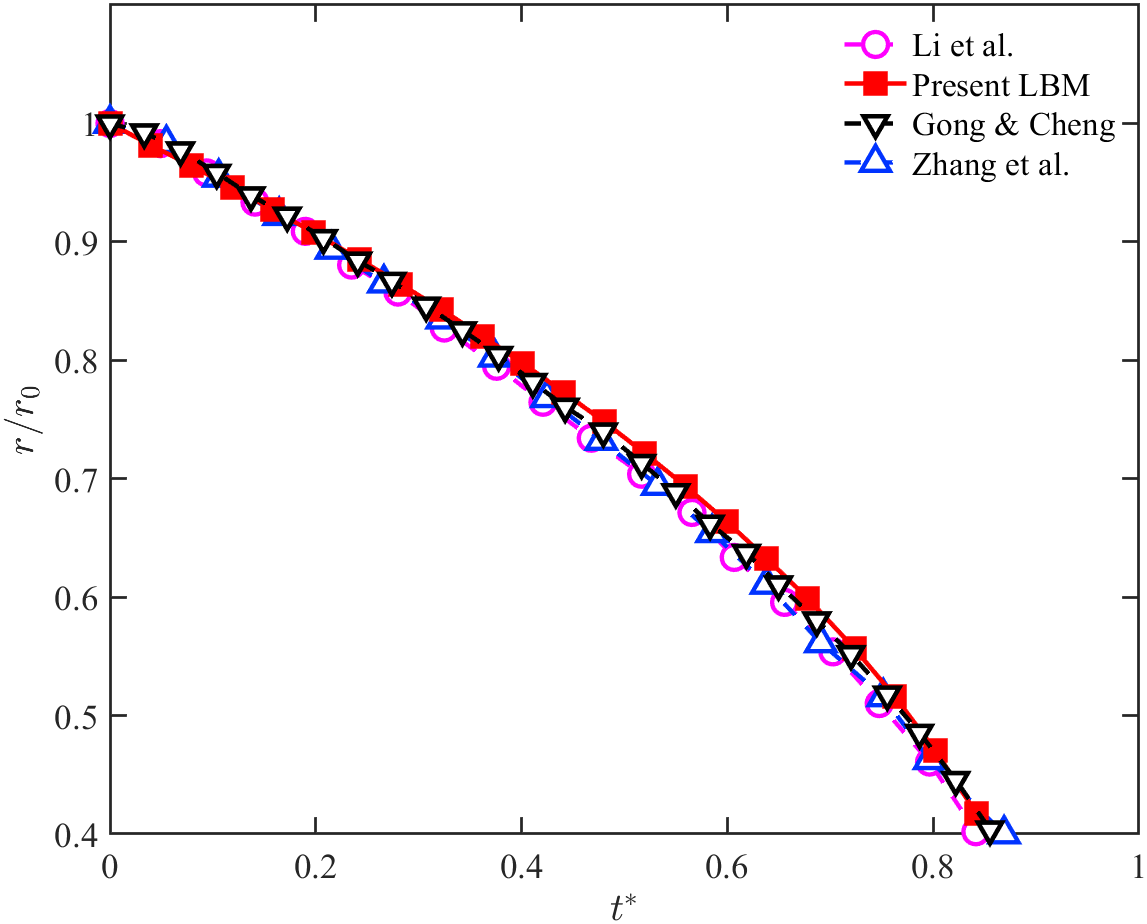}
   	\caption{Droplet evaporation on a solid surface.}
   	\label{fig:heated surface_nihe}
   \end{figure}

    The density contours obtained from the present simulations are shown in Figure~\ref{fig:heated surface}. To further validate the present model, the variations of the normalized contact radius obtained from different models are also presented in Figure~\ref{fig:heated surface_nihe}. Due to the diffusive nature of the interface in the LB model, the droplet radius at the final evaporation stage is excluded to ensure the accuracy of the tracked interface position. As shown in Figure~\ref{fig:heated surface_nihe}, the results from the present LB model and the previous LB models of Li et al.~\cite{Li2017}, Gong~\cite{Gong2012}, and Zhang et al.~\cite{Zhang2021} are in good agreement with the experimental data. It should be noted that, at the beginning of evaporation, the dimensionless radius $r/r_0$ exhibits a sharp transition due to the initial condition effect. Consequently, the initial radius $R_0$ is taken as the droplet radius after the initial transient, and the dimensionless time $t^*$ is referenced to the onset of evaporation. This treatment is the same as that of Zhang et al.~\cite{Zhang2021}.
   
\begin{figure}[htbp]
   	\centering
   	\includegraphics[width=0.4\textwidth]{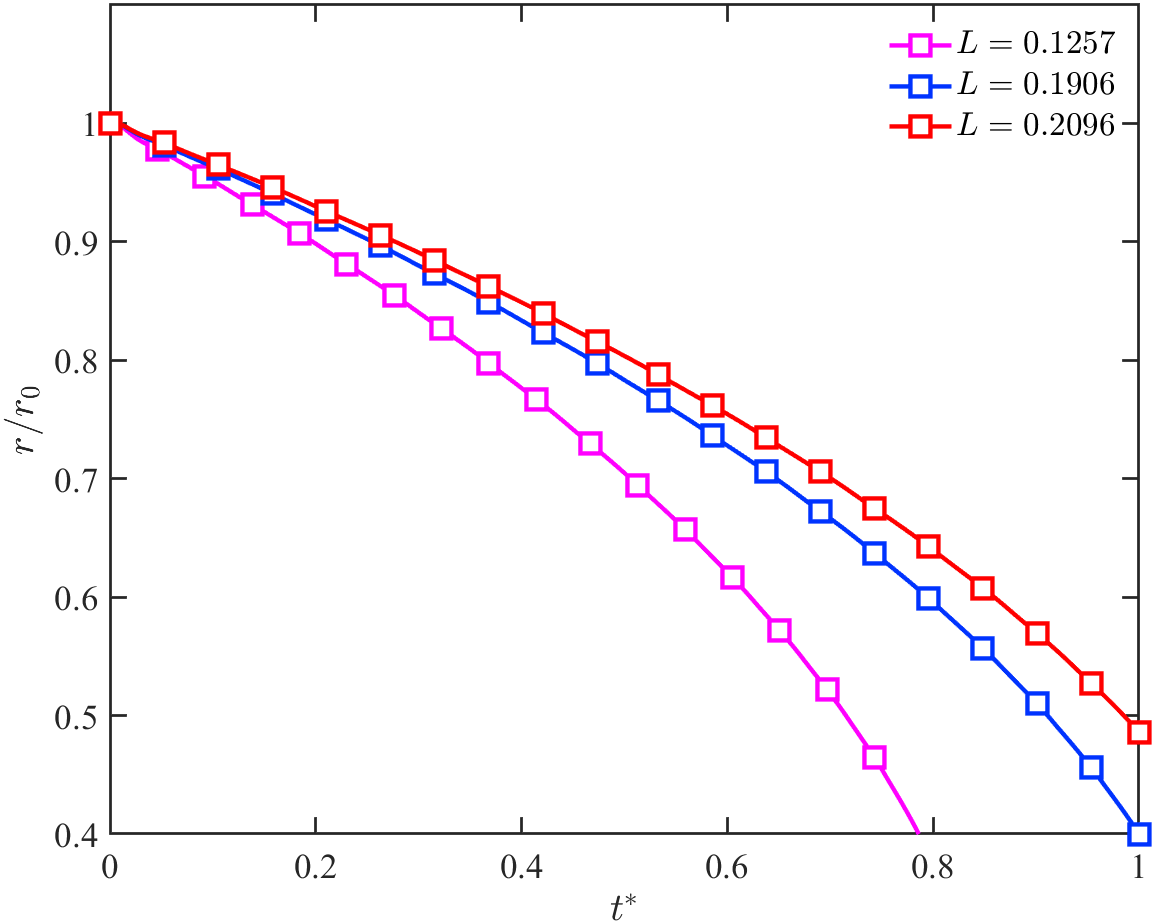}
   	\caption{Evolution of the dimensionless radius \(r/r_0\) with dimensionless time \(t^*\) for sessile-droplet evaporation at different latent heats.}
   	\label{fig:heated surface_Lduibi}
   \end{figure}

   Figure~\ref{fig:heated surface_Lduibi} shows the evolution of the dimensionless droplet radius \(r/r_0\) on the heated surface for different latent heats. It can be seen that the droplet radius decreases continuously with time for all the considered cases, while the evaporation rate decreases noticeably with increasing latent heat. For a smaller latent heat, less energy is required for liquid-vapor phase transition, resulting in a faster reduction of the droplet size. In contrast, increasing \(L\) significantly delays the evaporation process. These results demonstrate that the present total-enthalpy LB model can consistently describe the influence of latent heat on sessile-droplet evaporation. For the case of \(L=0.2096\), the late-stage evolution is increasingly influenced by the finite thickness of the diffuse interface. Since the characteristic droplet size becomes less well separated from the interfacial length scale, these late-stage data are not used for quantitative comparison.
   
   \subsection{Bubble nucleation and departure}
   In this subsection, to validate the present LB model, two-dimensional and three-dimensional numerical simulations are conducted for bubble nucleation and departure during nucleate boiling.
    
   \subsubsection{Two-dimensional simulation of bubble nucleation and departure}
      \begin{figure}[htbp]
   	\centering
   	\begin{minipage}{0.5\textwidth}  % 左右各留4%，图整体更大
   		\centering
   		\begin{subfigure}[b]{0.3\textwidth}   % 子图宽度增大
   			\centering
   			\includegraphics[trim=6cm 2cm 22cm 12cm, clip=true, width=\textwidth]{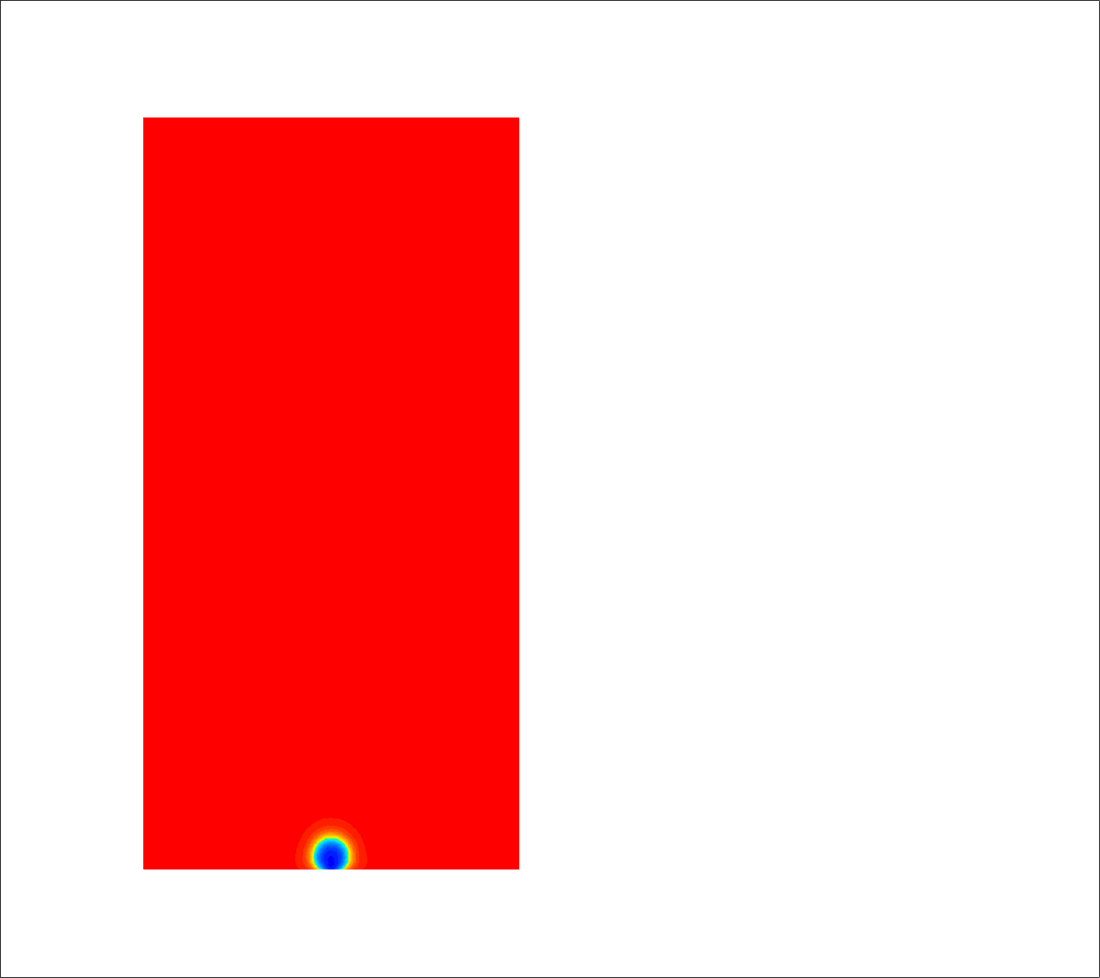}
   			\caption*{(a)}
   			\label{fig:Bubble_nucleation_Lsub1}
   		\end{subfigure}%
   		\hspace{0.01\textwidth}%              % 间距极小（0.5%）
   		\begin{subfigure}[b]{0.3\textwidth}
   			\centering
   			\includegraphics[trim=6cm 2cm 22cm 12cm, clip=true, width=\textwidth]{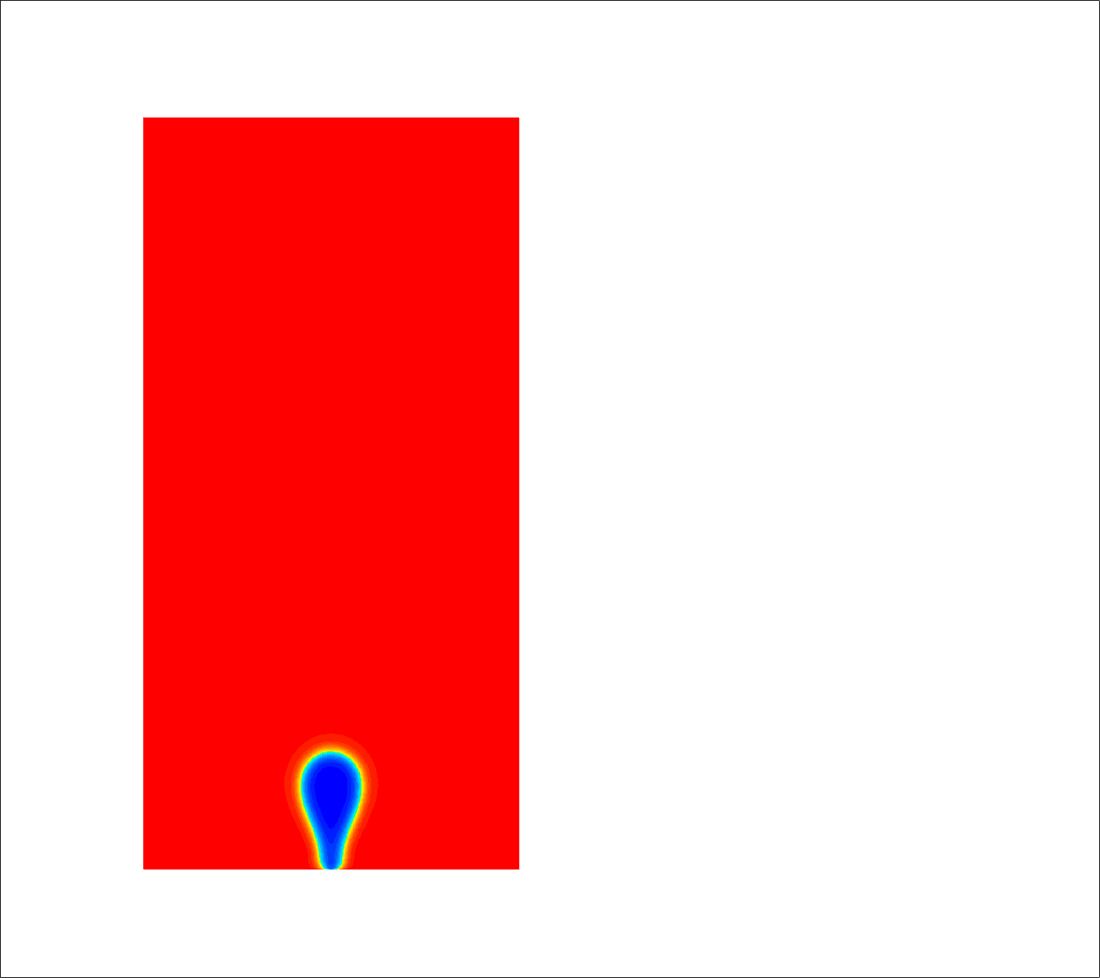}
   			\caption*{(b)}
   			\label{fig:Bubble_nucleation_Lsub2}
   		\end{subfigure}%
   		\hspace{0.01\textwidth}%              % 间距极小（0.5%）
   		\begin{subfigure}[b]{0.3\textwidth}
   			\centering
   			\includegraphics[trim=6cm 2cm 22cm 12cm, clip=true, width=\textwidth]{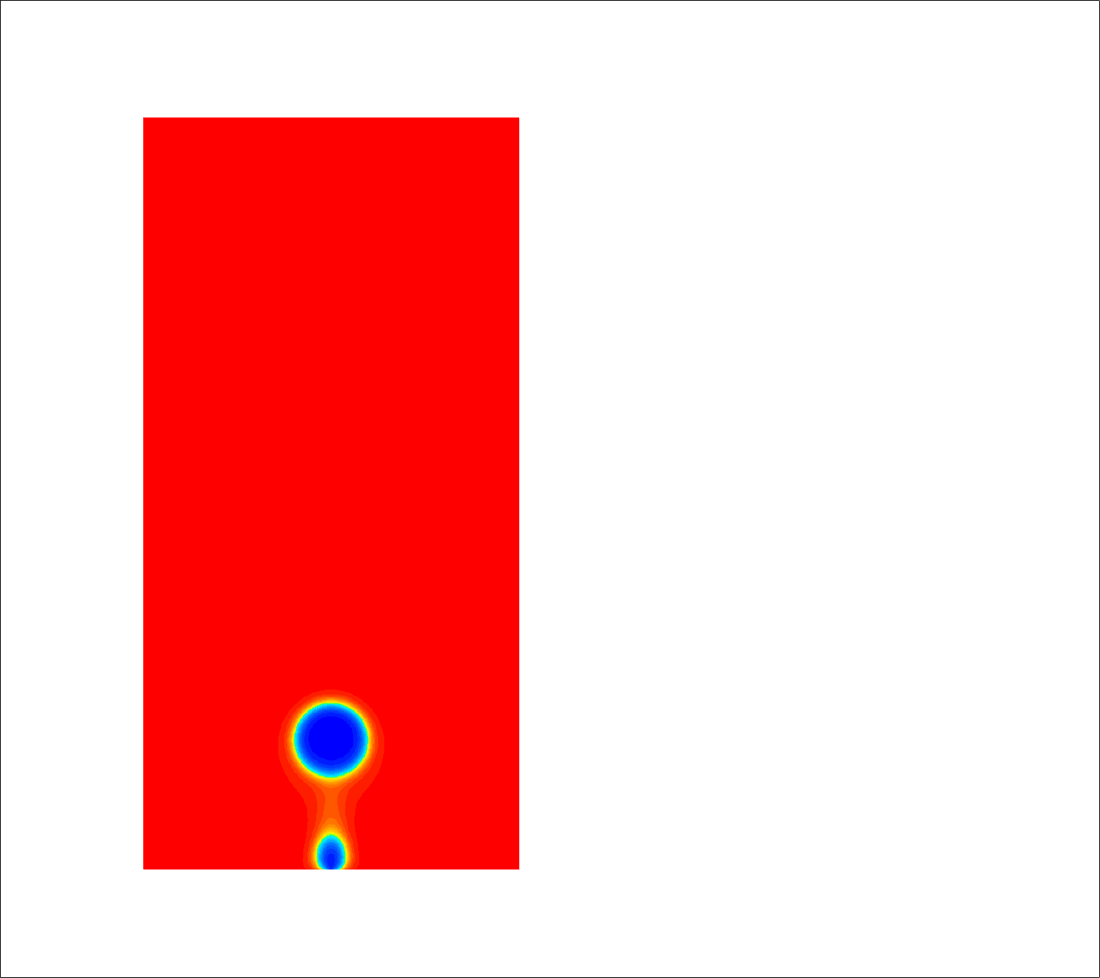}
   			\caption*{(c)}
   			\label{fig:Bubble_nucleation_Lsub3}
   		\end{subfigure}
   	\end{minipage}
   	\caption{Distributions of the density contours in nucleate boiling at $t = 4000$ (a), $t = 16000$ (b), and $t = 19000$ (c) ($g = 2.75 \times 10^{-5},L=0.1257$).}
   	\label{fig:Bubble_nucleation_L=0.1092}
   \end{figure}
   
   \begin{figure}[htbp]
   	\centering
   	\begin{minipage}{0.5\textwidth}  % 左右各留4%，图整体更大
   		\centering
   		\begin{subfigure}[b]{0.3\textwidth}   % 子图宽度增大
   			\centering
   			\includegraphics[trim=6cm 2cm 22cm 12cm, clip=true, width=\textwidth]{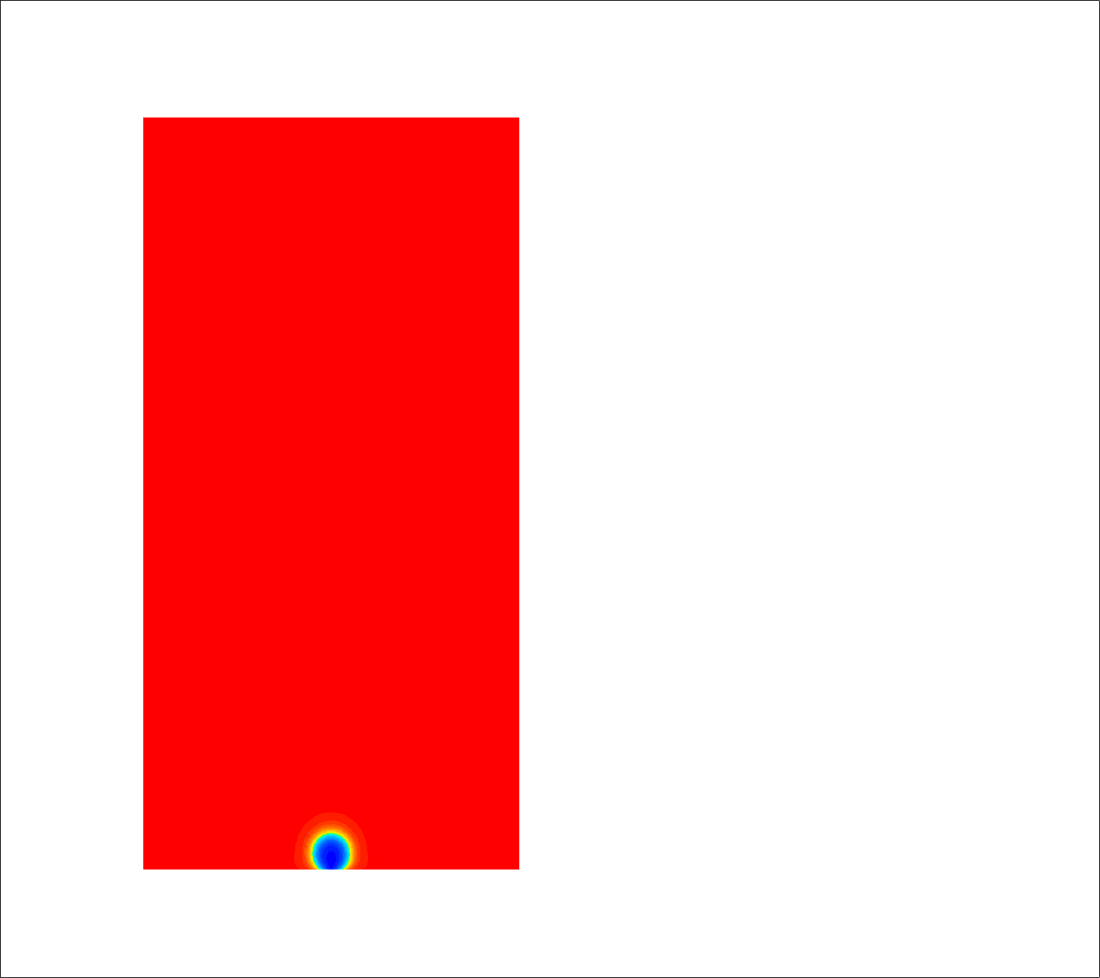}
   			\caption*{(a)}
   			\label{fig:Bubble_nucleation_sub1}
   		\end{subfigure}%
   		\hspace{0.01\textwidth}%              % 间距极小（0.5%）
   		\begin{subfigure}[b]{0.3\textwidth}
   			\centering
   			\includegraphics[trim=6cm 2cm 22cm 12cm, clip=true, width=\textwidth]{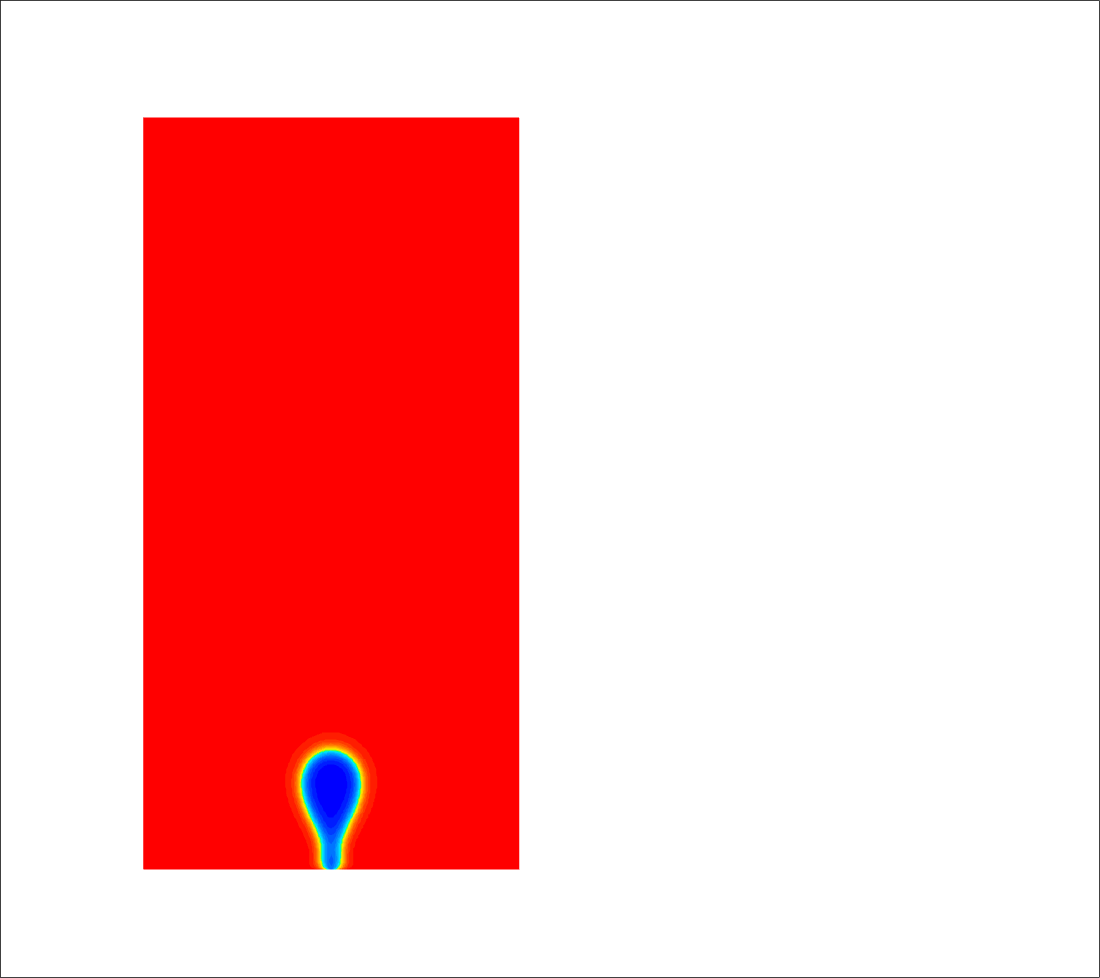}
   			\caption*{(b)}
   			\label{fig:Bubble_nucleation_sub2}
   		\end{subfigure}%
   		\hspace{0.01\textwidth}%              % 间距极小（0.5%）
   		\begin{subfigure}[b]{0.3\textwidth}
   			\centering
   			\includegraphics[trim=6cm 2cm 22cm 12cm, clip=true, width=\textwidth]{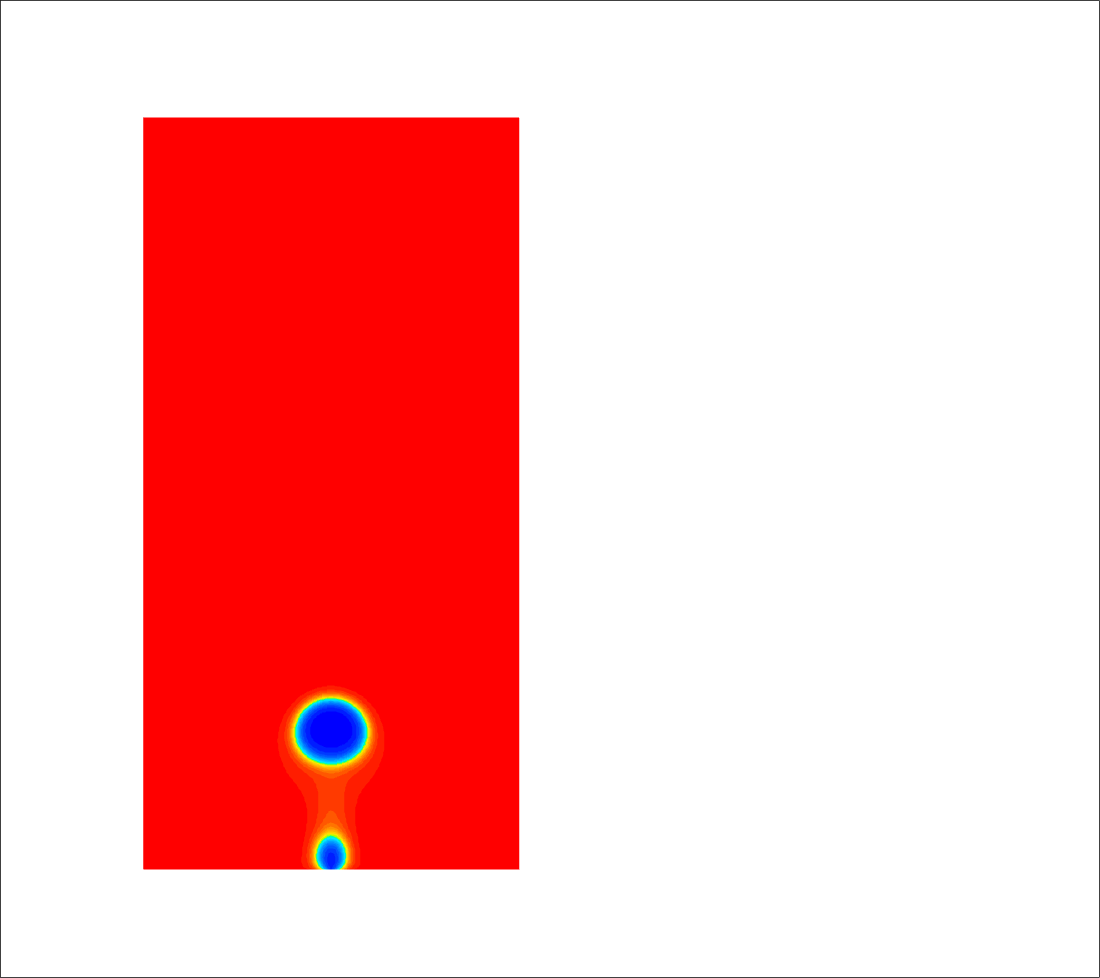}
   			\caption*{(c)}
   			\label{fig:Bubble_nucleation_sub3}
   		\end{subfigure}
   	\end{minipage}
   	\caption{Distributions of the density contours in nucleate boiling at $t = 6000$ (a), $t = 19000$ (b), and $t = 22000$ (c) ($g = 2.75 \times 10^{-5},L=0.1906$).}
   	\label{fig:Bubble_nucleation}
   \end{figure}
   We employ a $150 \times 300$ rectangular computational domain with periodic boundary conditions imposed along the horizontal direction. The computational domain is initially filled with saturated water, with the temperature uniformly initialized at $T_{sat}=0.86 T_c$. The velocity field is subject to the no-slip condition at the bottom solid wall and the open boundary condition at the top. A localized heating of $T_h=1.25 T_c$ is applied to the central five grid points on the bottom wall, while the remaining bottom wall and the top boundary are both held at the saturation temperature $T_{sat}$. Apart from the fluid-field physical parameters, which are the same as in the previous cases, and the contact angle at the solid surface is set to $45^{\circ}$. A vertical buoyant force $F_b=(\rho-\rho_{ave})g$ is also incorporated, with $\rho_{ave}$ being the domain averaged density and $g=(0,-g)$ the gravitational acceleration. The latent heat of vaporization is the same as in the Validation of the $D^2$ law.

   \begin{figure}[htbp]
   	\centering
   	\begin{subfigure}{0.5\linewidth}
   		\centering
   		\includegraphics[width=0.95\linewidth]{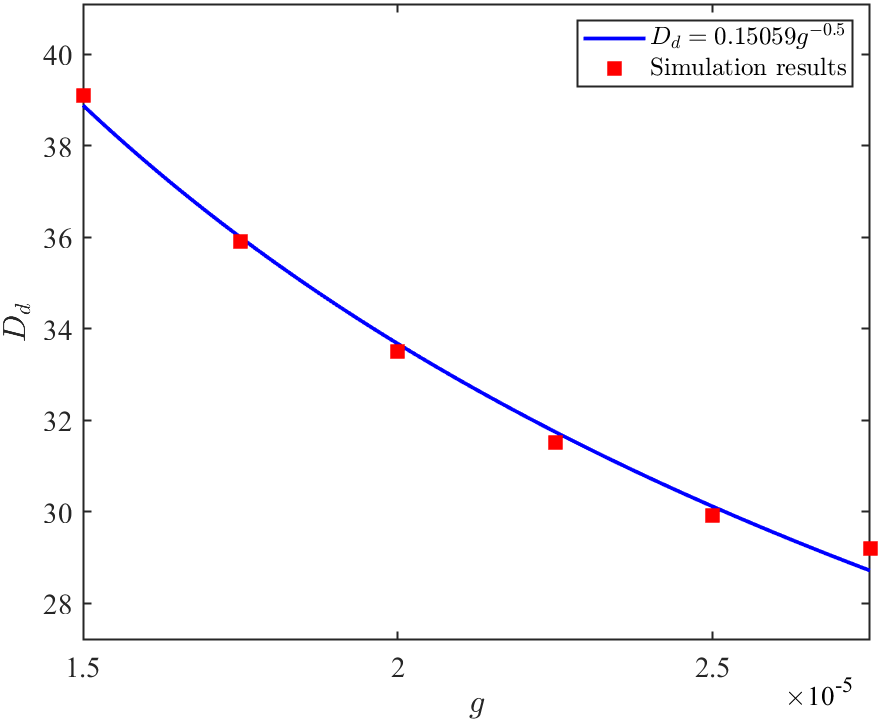}
   		\caption*{(a)}
   		\label{fig:bubble_a}
   	\end{subfigure}
   	
   	\begin{subfigure}{0.5\linewidth}
   		\centering
   		\includegraphics[width=0.95\linewidth]{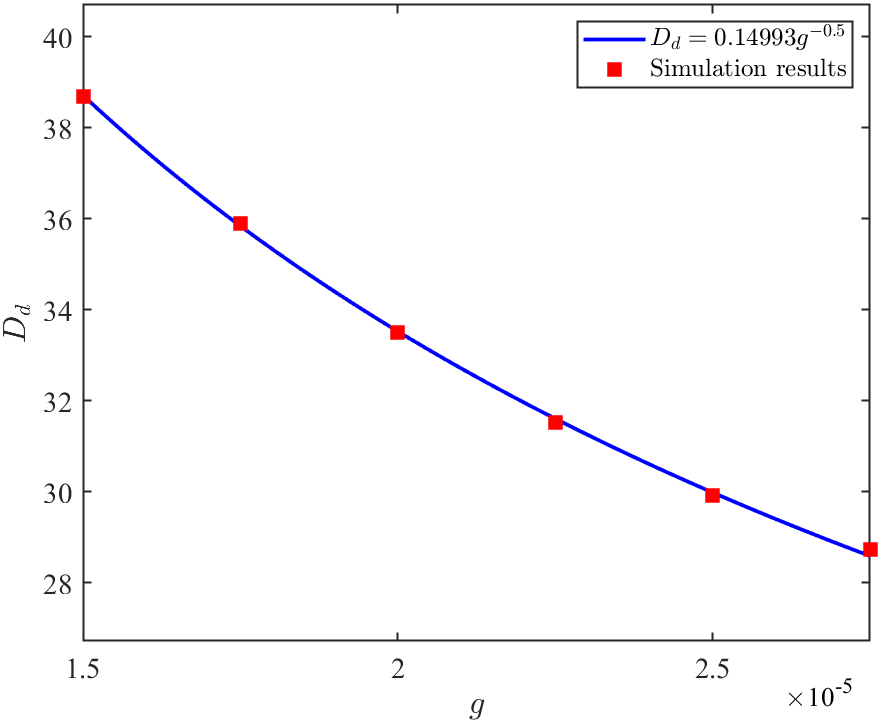}
   		\caption*{(b)}
   		\label{fig:bubble_b}
   	\end{subfigure}
   	
   	\caption{Simulation results for bubble nucleation and departure:
   		(a) $L=0.1257$; (b) $L=0.1906$.}
   	\label{fig:Bubble_nucleation_nihe}
   	
   \end{figure} 
    
   Figure~\ref{fig:Bubble_nucleation_L=0.1092} and Figure~\ref{fig:Bubble_nucleation} shows representative snapshots of nucleate boiling for two different latent heats. In both cases, a bubble nucleates at the heated wall, grows, detaches, and rises under buoyancy. A smaller latent heat $L=0.1257$ results in faster bubble growth and earlier departure, whereas increasing \(L=0.1906\) delays the phase-change process and bubble detachment. This comparison further demonstrates the capability of the present model to describe the effect of latent heat on boiling dynamics. And these processes of bubble growth and detachment observed from Figure~\ref{fig:Bubble_nucleation} and Figure~\ref{fig:Bubble_nucleation_L=0.1092} agree well with those described in the previous works~\cite{Zhang2021,Li2017,Gong2012}.
   
  \begin{figure*}[!t]%
  	\centering
  	% 第一行
  	\hspace*{0.05\textwidth}  % 左侧留白
  	\begin{minipage}[b]{0.2\textwidth}
  		\centering
  		\includegraphics[trim=0cm 0.2cm 0cm 0cm, clip=true, width=0.75\textwidth]{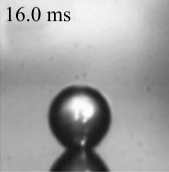}
  	\end{minipage}
  	\hspace*{0.005\textwidth}
  	\begin{minipage}[b]{0.2\textwidth}
  		\centering
  		\includegraphics[trim=0cm 0.2cm 0cm 0cm, clip=true, width=0.75\textwidth]{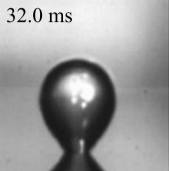}
  	\end{minipage}
  	\hspace*{0.005\textwidth}
  	\begin{minipage}[b]{0.2\textwidth}
  		\centering
  		\includegraphics[trim=0cm 0.2cm 0cm 0cm, clip=true, width=0.75\textwidth]{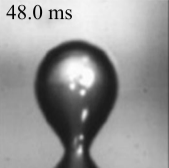}
  	\end{minipage}
  	\hspace*{0.005\textwidth}
  	\begin{minipage}[b]{0.2\textwidth}
  		\centering
  		\includegraphics[trim=0cm 0.2cm 0cm 0cm, clip=true, width=0.75\textwidth]{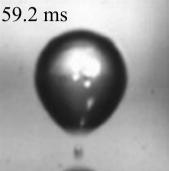}
  	\end{minipage}
  	\hspace*{0.08\textwidth}  % 右侧留白
  	\vspace{0.008\textheight} 
  	\textbf{(a)}
  	
  	% 第二行
  	\hspace*{0.03\textwidth}  % 左侧留白
  	\begin{minipage}[b]{0.22\textwidth}
  		\centering
  		\includegraphics[trim=7cm 2cm 9cm 12.5cm, clip=true, width=0.75\textwidth]{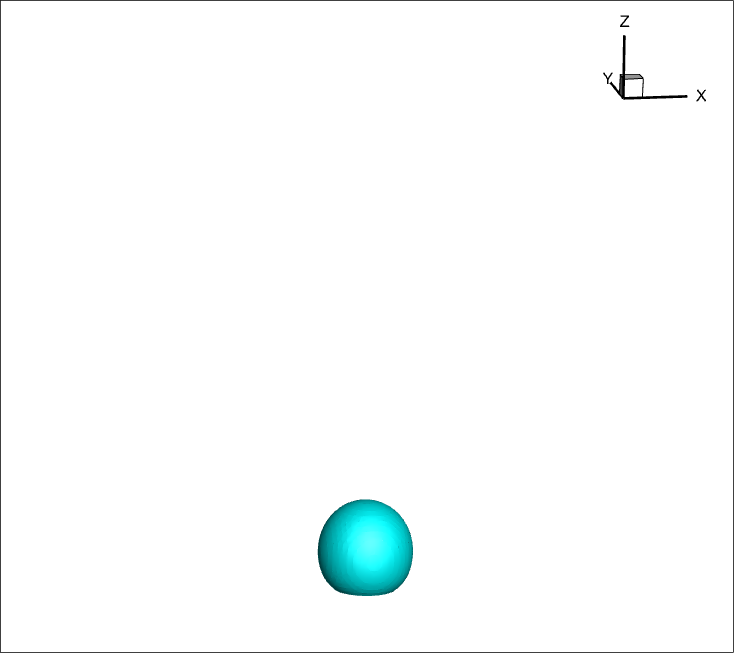}
  	\end{minipage}
  	\hspace*{0.002\textwidth}
  	\begin{minipage}[b]{0.22\textwidth}
  		\centering
  		\includegraphics[trim=8cm 2cm 8cm 12.5cm, clip=true, width=0.75\textwidth]{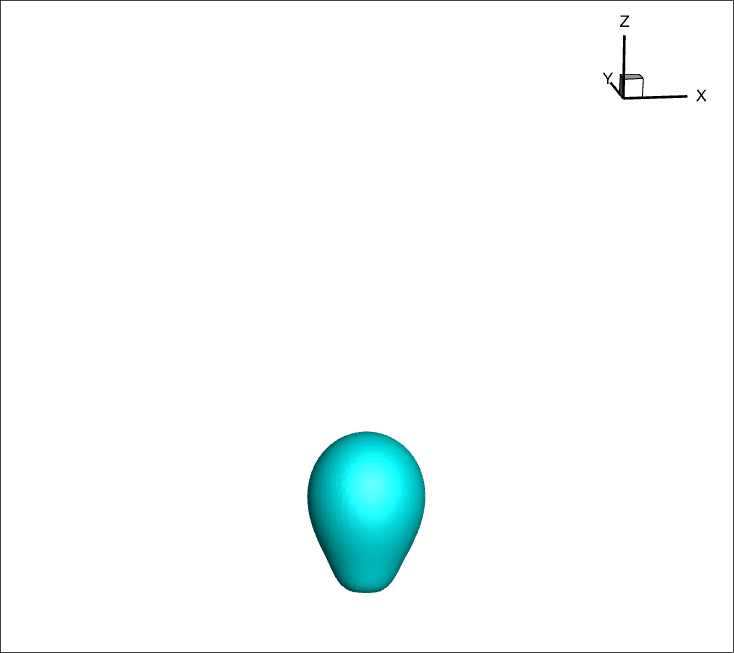}
  	\end{minipage}
  	\hspace*{0.005\textwidth}
  	\begin{minipage}[b]{0.22\textwidth}
  		\centering
  		\includegraphics[trim=10cm 2cm 8cm 12.5cm, clip=true, width=0.75\textwidth]{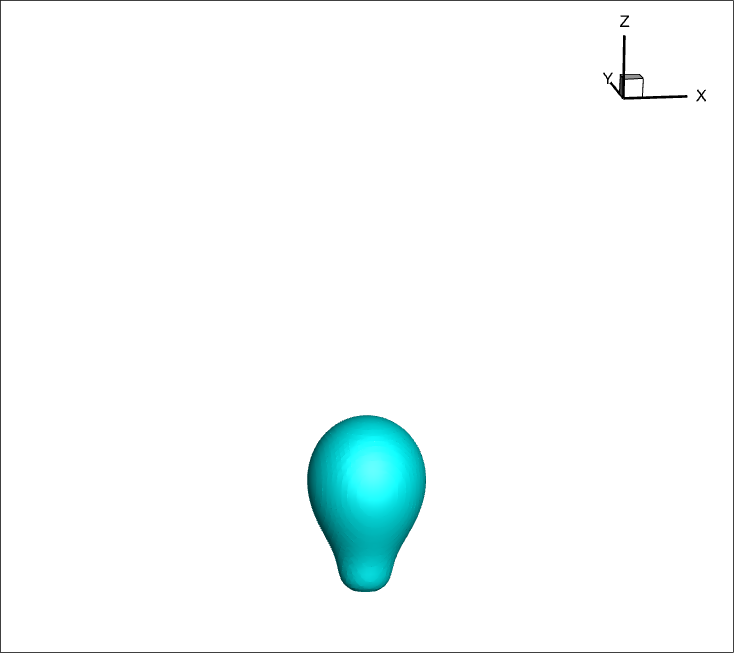}
  	\end{minipage}
  	\hspace*{0.005\textwidth}
  	\begin{minipage}[b]{0.22\textwidth}
  		\centering
  		\includegraphics[trim=10cm 2cm 7cm 12.5cm, clip=true, width=0.75\textwidth]{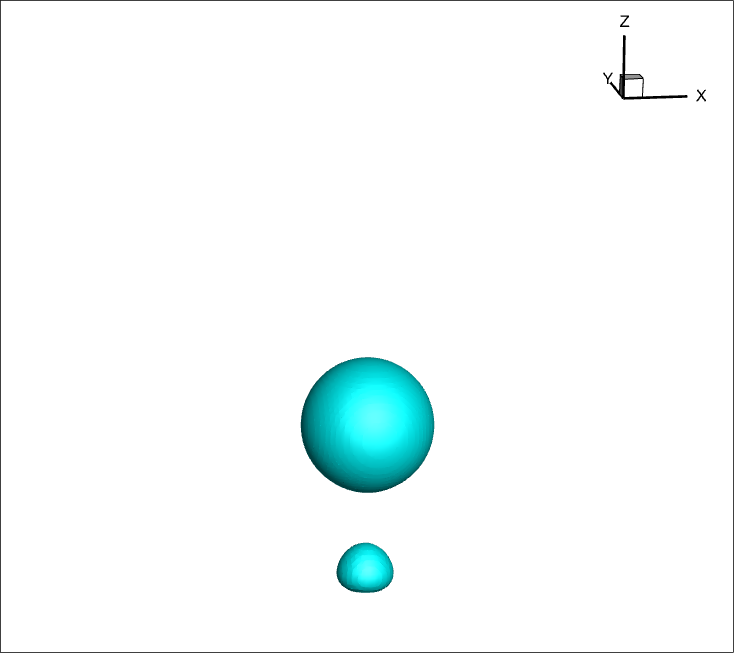}
  	\end{minipage}
  	\hspace*{0.03\textwidth}  % 右侧留白
  	\vspace{0.005\textheight} 
  	\textbf{(b)}
  	
  	% 第三行
  	\hspace*{0.05\textwidth}  % 左侧留白
  	\begin{minipage}[b]{0.2\textwidth}
  		\centering
  		\includegraphics[trim=0cm 0cm 0cm 0cm, clip=true, width=0.75\textwidth]{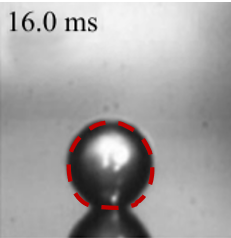}
  	\end{minipage}
  	\hspace*{0.005\textwidth}
  	\begin{minipage}[b]{0.2\textwidth}
  		\centering
  		\includegraphics[trim=0cm 0cm 0cm 0cm, clip=true, width=0.75\textwidth]{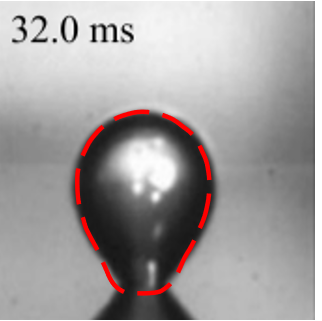}
  	\end{minipage}
  	\hspace*{0.005\textwidth}
  	\begin{minipage}[b]{0.2\textwidth}
  		\centering
  		\includegraphics[trim=0cm 0cm 0cm 0cm, clip=true, width=0.75\textwidth]{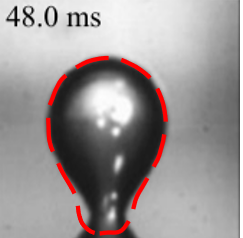}
  	\end{minipage}
  	\hspace*{0.005\textwidth}
  	\begin{minipage}[b]{0.2\textwidth}
  		\centering
  		\includegraphics[trim=0cm 0cm 0cm 0cm, clip=true, width=0.75\textwidth]{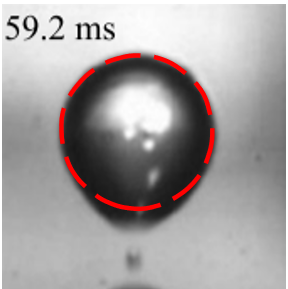}
  	\end{minipage}
  	\hspace*{0.08\textwidth}  % 右侧留白
  	\vspace{0.005\textheight} 
  	\textbf{(c)}
  	
  	% 第三行
  	\hspace*{0.05\textwidth}  % 左侧留白
  	\begin{minipage}[b]{0.2\textwidth}
  		\centering
  		\includegraphics[trim=0cm 0cm 0cm 0cm, clip=true, width=0.75\textwidth]{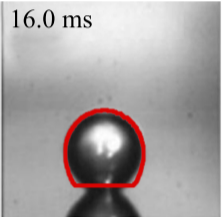}
  	\end{minipage}
  	\hspace*{0.005\textwidth}
  	\begin{minipage}[b]{0.2\textwidth}
  		\centering
  		\includegraphics[trim=0cm 0cm 0cm 0cm, clip=true, width=0.75\textwidth]{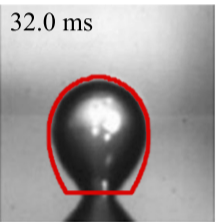}
  	\end{minipage}
  	\hspace*{0.005\textwidth}
  	\begin{minipage}[b]{0.2\textwidth}
  		\centering
  		\includegraphics[trim=0cm 0cm 0cm 0cm, clip=true, width=0.75\textwidth]{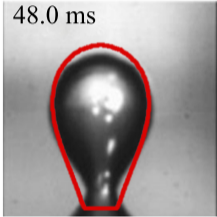}
  	\end{minipage}
  	\hspace*{0.005\textwidth}
  	\begin{minipage}[b]{0.2\textwidth}
  		\centering
  		\includegraphics[trim=0cm 0cm 0cm 0cm, clip=true, width=0.75\textwidth]{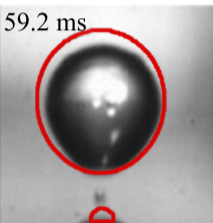}
  	\end{minipage}
  	\hspace*{0.08\textwidth}  % 右侧留白
  	\vspace{0.005\textheight} 
  	\textbf{(d)}
  	\caption{Bubble nucleation and departure: (a) experiment~\cite{Mukherjee2004}; (b) present simulation; (c) present simulation compared with experiment; (d) Qiao et al.~\cite{Qiao2026} compared with experiment.}
  	\label{fig:Bubble_nucleation_nihe_PR}
  \end{figure*}
For a more quantitative assessment, the departure bubble diameter is compared with the classical scaling law proposed by Fritz~\cite{Fritz1935}, which predicts \(D_d \propto g^{-0.5}\). A series of simulations is performed under different gravitational accelerations for two latent heats, and the results are summarized in Fig.~\ref{fig:Bubble_nucleation_nihe}. For both latent heats, the computed departure diameters closely follow the \(g^{-0.5}\) scaling. Although the absolute departure diameter varies slightly with \(L\), the dependence on gravitational acceleration remains essentially unchanged. These results demonstrate that the present model can reproduce the expected gravitational scaling of bubble departure over different latent heat conditions.

\subsubsection{Three-dimensional simulation of bubble nucleation and departure}

\begin{comment}
 We employ a $64 \times 64 \times 128$ cuboid computational domain with periodic boundary conditions imposed along the x-and y-directions. The computational domain is initially filled with saturated water, with the temperature uniformly initialized at $T_{sat}=0.86 T_c$. The velocity field is subject to the no-slip condition at the bottom solid wall and the open boundary condition at the top. A localized heating of $T_h=1.25 T_c$ is applied to the central seven grid points on the bottom wall, while the remaining bottom wall and the top boundary are both held at the saturation temperature $T_{sat}$. Apart from the fluid-field physical parameters, which are the same as in the previous cases, and the contact angle at the solid surface is set to $45^{\circ}$. A vertical buoyant force $F_b=(\rho-\rho_{ave})g$ is also incorporated, with $\rho_{ave}$ being the domain averaged density and $g=(0,0,-g)$ the gravitational acceleration. The latent heat of vaporization is the same as in the Validation of the $D^2$ law.
\end{comment}

 For the three-dimensional simulation, a $64 \times 64 \times 128$ cuboid computational domain is employed, with periodic boundary conditions imposed in the $x$- and $y$-directions. The initial and thermal boundary conditions are otherwise identical to those of the two-dimensional case, except that the localized heating region covers the central seven grid points on the bottom wall. The contact angle is set to $45^{\circ}$, the thermal conductivity is \(\lambda=1/3\) and the gravitational acceleration is prescribed as $\mathbf{g}=(0,0,-g)$. The same latent heat of vaporization as that used in the $D^2$-law validation is adopted.

Fig.~\ref{fig:Bubble_nucleation_nihe_PR} compares the bubble evolution predicted by the present thermal lattice Boltzmann model with the experimental observations reported by Mukherjee and Dhir~\cite{Mukherjee2004}. The numerical results reproduce the main stages of bubble nucleation, growth, necking, and departure, and show good qualitative agreement with the experimentally observed bubble morphology and evolution.Compared with the results reported by Qiao et al.~\cite{Qiao2026}, the present model appears to reproduce the experimentally observed bubble morphology more closely, particularly during the necking and departure stages.

\begin{figure}[htbp]
	\centering
	\includegraphics[width=0.4\textwidth]{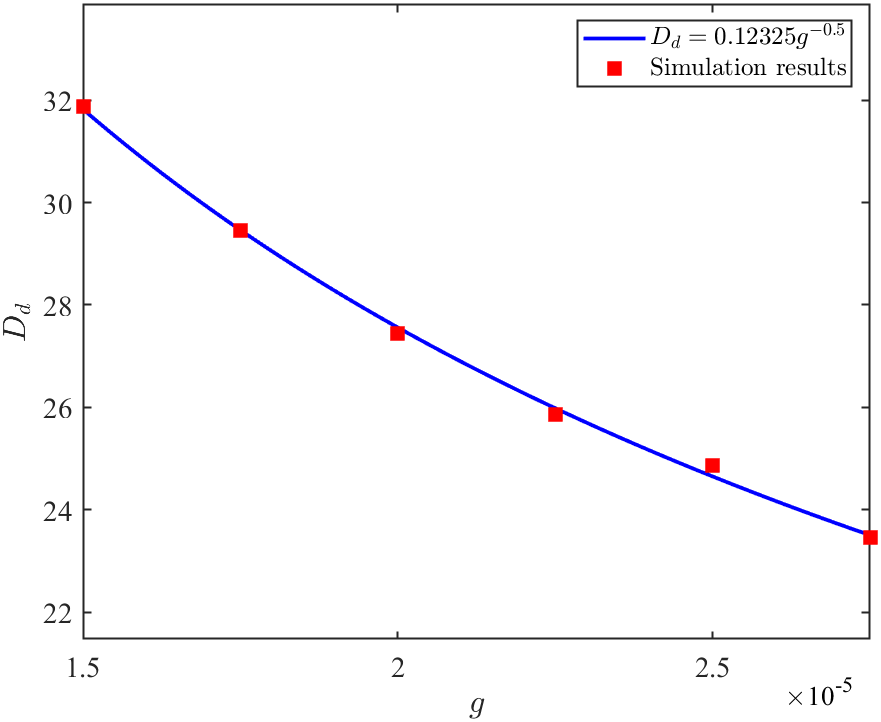}
	\caption{Simulation results for bubble nucleation and departure.}
	\label{fig:Bubble_nucleation_duibi_PR}
\end{figure}

To further assess the model in three dimensions, the dependence of the
departure bubble diameter on gravitational acceleration is examined using the same procedure as in the two-dimensional case. As shown in
Fig.~\ref{fig:Bubble_nucleation_duibi_PR}, the computed departure diameters remain in good agreement with the Fritz scaling. This result indicates that the present model preserves the expected gravitational dependence of bubble departure when extended to three-dimensional nucleate boiling simulations.

   \section{CONCLUSION}
   	\label{sec:CONCLUSION}
   %\vspace{1.8em}
   	In this work, an MRT total enthalpy LB model is developed for liquid-vapor phase change heat transfer. Multiscale analysis confirms that the target macroscopic energy equation can be recovered without introducing the nonlinear thermodynamic source term commonly required in temperature-based formulations,thereby avoiding the explicit evaluation of $\nabla\cdot\mathbf{u}$.
   	The numerical results reproduce the characteristic behavior of droplet evaporation and bubble growth and departure, including the $D^2$ law and the Fritz scaling, with good agreement with published results. Meanwhile, the latent heat can be specified independently in the thermal model, allowing the caloric properties to be adjusted according to the physical problem considered, while the separately derived EOS-based expression provides a thermodynamic reference for its selection.
   	
   \begin{acknowledgments}
   	 The authors thank Prof. Rongzong Huang for his invaluable assistance with discussions on theoretical modeling.
   	 
   	This work was  supported by the following grants: the National Natural Science Foundation of China (Grant Nos. 12302373, 12472297 and 12372286 ),  the Outstanding Youth Project of the Hunan Provincial Education Department (Grant No. 24B0310), the Science and Technology Innovation Program of Hunan Province (Grant No. 2024RC3228) , and the Research Foundation of Education Bureau of Hunan Province (Grant No. 25A0666).
   \end{acknowledgments}

	\bibliography{document}
	
\end{document}